# ESTRUCTURA, BIOMASA AÉREA Y CARBONO ALMACENADO EN LOS BOSQUES DEL SUR Y NOROCCIDENTE DE CÓRDOBA

Adela Vásquez & Henry Arellano

**RESUMEN**

Se estimó la biomasa aérea y el carbono almacenado en doce tipos de bosque en localidades del departamento de Córdoba, cuyos montos anuales de precipitación varían desde 3000 mm (climas súper húmedos) hasta 1300 mm (climas semihúmedos). La biomasa se estimó con base en aspectos de la estructura de la vegetación (diámetro a la altura del pecho, altura y peso específico de la madera de los individuos). Para estimar la biomasa se probaron nueve ecuaciones alométricas generadas para bosques tropicales disponibles en la literatura y se seleccionaron las ecuaciones propuestas por Chave *et al.* (2005) específicas para bosques húmedos y secos. En levantamientos tipo, se midió el contenido de carbono de los árboles en cuatro tejidos (tallo, rama, corteza y hojas), mediante el método de combustión en seco en un analizador automático que permite obtener el porcentaje de carbono total presente en una muestra de peso conocido. Para cuantificar el carbono almacenado en la biomasa, se multiplicaron estas fracciones de carbono obtenidas por unidad de masa seca y la biomasa estimada para cada tejido; a partir de la suma de cada tejido se obtuvo el carbono almacenado por individuo. Los valores de biomasa y carbono más altos se obtuvieron en las asociaciones Prestoeo decurrentis - Trichillietum poeppigi, Macrolobio ischnocalycis - Peltogynetum purpurea con valores de biomasa que oscilan entre 40.83 y 41.15 Mg/0.05 ha y de carbono entre 18.44 y 18.9 MgC/0.05ha, seguidos por los bosques de las asociaciones Protio aracouchini - Viroletum elongatae y Tovomito weddellianae - Quercetum humboldtii, con valores altos de biomasa entre 30.01 y 34.33 Mg/0.05 ha y carbono entre 13.78 y 15.76 MgC/0.05 ha. Estos resultados se relacionan con valores altos de área basal y altura de los individuos de estos bosques. Los valores más bajos se obtuvieron en los bosques de asociaciones Marilo laxiflorae - Pentaclethretum macrolobae, Mayno grandifoliae - Astrocaryetum malybo, Cordietum proctato - panamensis y la comunidad de *Acalypha* sp. y *Guazuma ulmifolia* cuya biomasa varía entre 7.39 y 14.09 Mg/0.05 ha y la cantidad carbono entre 3.34 y 6.58 MgC/0.05 ha. Estos resultados se relacionan con los valores bajos de área basal y altura, así como la tendencia a tener un alto número de individuos que se observa en estos bosques. En la mayoría de los bosques la biomasa y el carbono se concentran principalmente en las clases diamétricas superiores (> 70cm) y en los estratos arbóreos (superior e inferior). Los bosques de las asociaciones Prestoeo decurrentis-Trichillietum poeppigi, Macrolobio ischnocalycis-Peltogynetum purpurea y Trichilio hirtae-Schizolobietum parahibi_1 presentan dicha situación lo que explica sus valores altos de biomasa y carbono. En contraste, la asociación Cordietum proctato - panamensis y la comunidad de *Acalypha* sp. y *Guazuma ulmifolia* tienen mayor cantidad de biomasa acumulada (57.91% y 62.71% respectivamente) en las clases diamétricas menores –clases II (10-20cm) y III (30-50cm)- y en el estrato subarbóreo, situación que se relaciona con sus valores bajos de biomasa y carbono. Con respecto al gradiente de precipitación, la biomasa y el carbono variaron desde 34.18 Mg/0.05 ha y 15.68 MgC/0.05 ha respectivamente, para los bosques en clima súper húmedo hasta 7.39 Mg/0.05 ha y 3.34 MgC/0.05 ha respectivamente, para los semihúmedos; los bosques en clima muy húmedo y húmedo presentaron valores similares de 21.15 Mg/0.05 ha de biomasa y 9.27 MgC/0.05 ha de carbono para los primeros y de 21.4 Mg/0.05 ha de biomasa y 9.64 MgC/0.05 ha de carbono para los últimos. Probablemente, además de la precipitación, en estos resultados influye el estado de conservación de los bosques, ya que la intervención es más pronunciada a medida que avanza la condición de clima muy húmedo a semihúmedo, que podría incidir en la alta variabilidad observada en los bosques de estos ambientes. La contribución de las especies a las reservas de biomasa y carbono se da de forma desigual, y se encontró que la mayor cantidad de carbono está almacenado en la biomasa de un reducido número de especies. Entre las especies con esta dominancia en la biomasa y carbono, se encuentran *Quercus humboldtii* que tiene almacenados 18.86 Mg/0.05 ha (62.99%) y 8.68 MgC/0.05 ha (62.93%); *Cavanillesia platanifolia* con 8.36 Mg/0.05 ha (34.74%) y 3.78 (34.88%) MgC/0.05 ha; *Peltogyne purpurea* con 15.57 Mg/0.05ha (37.88%) y 7.11 MgC/0.05ha (37.61%); *Dipteryx oleífera* con 7.97 Mg/0.05 ha (23.25%) y 3.76 MgC/0.05 ha (23.99%) de biomasa y carbono respectivamente.

**ABSTRACT**

We estimated the aerial biomass and stored carbon in twelve types of forests in the department of Córdoba with annual rain fall ranging from 3000 mm (super humid climates) to 1300 mm (semi-humid climate). Biomass was estimated based on structural aspects of the vegetation (diameter at breast height, total height, and wood specific weight). We tested nine allometric equations for tropical forests available in the literature and selected those proposed by Chave *et al* (2005) that are specific for humid and dry forests. Carbon content in trees was measured in four tissues (stem, branch, bark, and leafs) through an automated dry combustion method, which estimates the percentage of carbon in a sample of known weight. To quantify the stored carbon in the biomass, such percentages of carbon per dry mass were multiplied by the biomass estimated for each tissue; the stored carbon for each individual was then estimated by adding the biomass value of each tissue. Highest biomass and carbon values were found in the associations Prestoeo decurrentis - Trichillietum poeppigi, Macrolobio ischnocalycis - Peltogynetum purpurea (biomass 40.83 - 41.15 Mg/0.05 ha; carbon 18.44- 18.9 MgC/0.05ha), followed by forests of the asociaciones Protio aracouchini - Viroletum elongatae and Tovomito weddellianae - Quercetum humboldtii (biomass 30.01-34.33 Mg/0.05 ha; carbon 13.78-15.76 MgC/0.05 ha). The lowest values were found in forests of the associations Marilo laxiflorae - Pentaclethretum macrolobae, Mayno grandifoliae - Astrocaryetum malybo, Cordietum proctato – panamensis, and the community *Acalypha* sp. and *Guazuma ulmifolia* (biomass 7.39-14.09 Mg/0.05 ha; carbon 3.34- 6.58 MgC/0.05 ha). Such results are related with the low values in basal area and height as well as the tendency of having a high number of individuals within these forests. In most forests, biomass and carbon are concentrated in the arboreal strata (superior and inferior) and in individuals with a diameter greater than 70 cm, such as the forests of the associations Prestoeo decurrentis-Trichillietum poeppigi, Macrolobio ischnocalycis-Peltogynetum purpurea, and Trichilio hirtae-Schizolobietum parahibi_1. In contrast, the association Cordietum proctato – panamensis and the community *Acalypha* sp. and *Guazuma ulmifolia* have the greatest accumulated biomass (57.91% and 62.71%, respectively) in the subarboreal stratum with individuals of smaller diameter (diametric Classes II and III with 10-20cm and 30-50cm in diameter), which is related with their low values in biomass and carbon. Biomass and carbon varied with precipitation, ranging from 34.18 Mg/0.05 ha and 15.68 MgC/0.05 ha, respectively, in super humid forests to 7.39 Mg/0.05 ha and 3.34 MgC/0.05 ha in semi-humid forests; similar values were found in very humid (biomass 21.15 Mg/0.05 ha; carbon 9.27 MgC/0.05 ha) to humid forests (biomass 21.4 Mg/0.05; carbon 9.64 MgC/0.05 ha). It is possible that in addition to precipitation, the conservation state of each forest also affects these values because disturbance increases with an increment in the climatic conditions, from the very humid to semi-humid. The species contribution to the biomass and carbon are not equal, with most carbon stored in the biomass of a reduced number of species. Among those species are *Quercus humboldtii* with 18.86 Mg/0.05 ha (62.99%) of biomass and 8.68 MgC/0.05 ha (62.93%) of carbon; *Cavanillesia platanifolia* with 8.36 Mg/0.05 ha (34.74%) of biomass and 3.78 (34.88%) MgC/0.05 ha of carbon; *Peltogyne purpurea* with15.57 Mg/0.05ha (37.88%) of biomass and 7.11 MgC/0.05ha (37.61%) of carbon; and *Dipteryx oleifera* with 7.97 Mg/0.05 ha (23.25%) of biomass and 3.76 MgC/0.05 ha (23.99%) of carbon.

**INTRODUCCIÓN**

Los bosques a través del proceso de fotosíntesis capturan dióxido de carbono atmosférico ($CO_2$), lo fijan en sus estructuras vivas y parte de éste lo acumulan en su biomasa, lo transfieren al mantillo en descomposición y al suelo; de esta manera

Vásquez & Arellanoconstituyen reservas de carbono. Otra parte del carbono es intercambiado con la atmósfera, mediante los procesos de respiración y disturbio (Gower, 2003). La acumulación de carbono es influenciada principalmente por factores físicos, edáficos y por patrones de disturbio que afectan la estructura comunitaria y las reservas de biomasa y carbono en los bosques tropicales. La variación de la estructura y de los patrones de distribución de biomasa en los bosques tropicales se asocian principalmente con gradientes latitudinales y altitudinales que se relacionan con las diferencias climáticas, con las características físicas y químicas de los suelos, las condiciones topográficas y con las condiciones de humedad del suelo (Golley, 1983; Clark & Clark, 2000; Alves *et al.*, 2010; Laumonier, 2010; Wittmann & Thaíz Zorzi, 2008). Los procesos de disturbio también se consideran como condicionantes en esta variación (Urquiza-Haas *et al.*, 2007). La combinación de estos factores influye en la conformación de unidades ecológicas con características propias.

Estimar las reservas de biomasa de los bosques es una herramienta útil para valorar la cantidad de carbono que se almacena en las estructuras vivas en un momento dado, lo cual es importante para evaluar su contribución al ciclo del carbono. De ahí el interés por realizar estimaciones de biomasa en los bosques tropicales (Brown, 1997).

En Colombia, se han realizado estimaciones de biomasa en bosques en distintas regiones naturales del país. Las regiones andina y amazónica cuentan con el mayor número de estimaciones, mientras que la región Caribe tiene muy pocos estudios de este tipo (Anzola & Rodríguez, 2001).

Los bosques del departamento de Córdoba han sido sometidos a fuertes procesos de disturbio como deforestación y fragmentación, procesos que inciden en la dinámica de acumulación de carbono en su biomasa, por tanto es necesario cuantificar las reservas de biomasa actuales de estos bosques; se trata de cuantificar el carbono almacenado que presentan y de esta manera contribuir en la definición de estrategias de conservación que garanticen el mantenimiento de las reservas actuales, así como la continuidad de los servicios ambientales. La conservación de estos bosques y su manejo sostenible constituyen una estrategia apropiada de mitigación del cambio climático, al reducir las emisiones de carbono producto de la deforestación y al permitir que se siga acumulando más carbono. Aunque en los bosques que todavía se conservan y los relictos boscosos del Sur y del Noroccidente del departamento de Córdoba se realizó recientemente la caracterización florística y estructural de la vegetación, no se cuenta aún con información acerca de la cantidad de biomasa almacenada en estos ecosistemas ni de los factores relacionados con su distribución. Por lo tanto, en el presente estudio se ha propuesto estimar la biomasa aérea de los bosques asociados al gradiente de precipitación en localidades del Sur y Noroccidente del departamento de Córdoba mediante metodologías tradicionales y resolver los siguientes interrogantes:

(1) ¿Cuál es el arreglo en la estructura de los bosques asociados al gradiente de precipitación en localidades del Sur y Noroccidente del departamento de Córdoba?
(2) ¿Cuánto representa la reserva de biomasa aérea y de carbono que contienen?
(3) ¿Cómo se distribuye la biomasa aérea y el carbono en estos bosques?
(4) ¿La distribución de la biomasa aérea se relaciona con la precipitación?

**METODOLOGÍA**

**Área de estudio**

El estudio se desarrolló en localidades del Sur y Noroccidente el departamento de Córdoba, éste se ubica al Noroccidente de Colombia entre los 09º 26' 16" y 07º 22' 05" de latitud Norte y 74º 47' 43" y 76º 30' 01'' de longitud Oeste, en la región Caribe.

**Características del área de estudio**

En la geografía de Córdoba se distinguen dos regiones: la cordillerana y la plana. La zona montañosa se encuentra al Sur del departamento en las estribaciones de la cordillera occidental, que se ramifica en tres serranías, Abibe, San Jerónimo y Ayapel. La zona plana formada por la gran llanura del Caribe con ligeras ondulaciones con altitudes inferiores a 100 m incluye los valles delos ríos Sinú, San Jorge y el área costera (García, 1982).

El Departamento de Córdoba se caracteriza por un régimen térmico isotermal, donde el promedio anual siempre es superior a 26.5ºC. La temperatura media anual del aire varía entre 26.5ºC y 28ºC, en la mayor extensión del territorio se registran valores anuales promedio de 27 y 27.5ºC; los valores mínimos se presentan en el sur del departamento, de 26.5 a 27ºC y los valores más altos se registran en Ciénaga de Oro, 27.8ºC. La variación de la temperatura es poco diferenciable de un municipio a otro (Palencia *et al.*, 2006).

El clima de Córdoba presenta un único patrón de lluvias de tipo bimodal tetraestacional (Rangel & Arellano, 2010). La precipitación, la humedad y la evapotranspiración potencial (ETP) presentan variaciones fuertes en el territorio cordobés. Córdoba presenta un gradiente de precipitación que aumenta de norte a Sur. En el sector norte se registran valores entre 1000-1500 mm mientras que en el sector Sur en la zona montañosa, los montos de precipitación superan los 3000 mm al año (Rangel & Arellano, 2010; Rangel & Carvajal, 2011).

Para la realización del presente estudio se tomaron dos sectores de muestreo, el Sur y el Noroccidente de Córdoba.

En el sector Sur se estudiaron bosques establecidos en los municipios de Montelíbano, Tierralta y Valencia. El primer municipio pertenece a la subregión San Jorge y los dos últimos a la subregión Sinú (IGAC, 2009). La precipitación media anual en estos municipios varía entre 1600 y 2900 mm, con valores mayores en el extremo Sur de Tierralta y Montelíbano. La humedad relativa presenta valores entre a 84-85%. La temperatura promedio es de 26.9 °C (Palencia *et al.*, 2006).

Una clasificación local realizada por Palencia *et al.* (2006) ubica las localidades de Montelíbano y Tierralta como súper-húmedo y Valencia como semi-húmedo. De acuerdo con la definición de las unidades climáticas para la región Caribe de Colombia, realizada recientemente por Rangel & Carvajal (2012) con base en los montos anuales de precipitación, las localidades de muestreo en Montelíbano y el Sur de Tierralta (Parque Nacional Natural Paramillo) pertenecen a la unidad climática súper-húmedo, las otras localidades de Tierralta y Valencia corresponden a la unidad muy húmedo (Tabla 164).

En el Noroccidente se estudiaron algunos relictos boscosos de localidades de Los Córdobas, Canalete, Moñitos y Puerto Escondido, municipios pertenecientes a la zona plana, subregión costanera. En estos municipios la precipitación media anual varía entre 1300 mm y 1480 mm. La humedad relativa oscila entre 83-84% y la temperatura alcanza valores hasta de 28ºC (Palencia *et al.*, 2006).

Los relictos boscosos de Los Córdobas y Canalete se encuentran en la transición de bosque húmedo-seco, la clasificación local realizada por Palencia *et al.* (2006) con base en Thornthwaite, el clima es semi-húmedo en Los Córdobas y semiseco en Canalete. Además, en caracterizaciones recientes realizadas en estas localidades, se han encontrado representantes propios de bosque húmedo que constituyen puntos de enlace con la biota y la vegetación de regiones más húmedas (Rangel *et al.*, 2010), que permite ubicarlos en la denominación de bosques en climas húmedos. De acuerdo con los montos de precipitación anual de las localidades de muestreo y con las unidades climáticas definidas por Rangel & Carvajal (2012), los relictos boscosos de Los Córdobas y Canalete pertenecen a la unidad de clima húmedo y los relictos boscosos de Moñitos y Puerto Escondido pertenecen a la formación de bosque semi-húmedo (Tabla 164).

Córdoba es un departamento donde los bosques han sido sometidos a altos procesos de transformación antrópica, dejando como resultado relictos boscosos en el norte y bosques con diferentes grados de fragmentación y conservación en el Sur. Los sitios muestreados incluyen bosques con bajo grado de intervención aparente hasta bosques con niveles muy altos (Tabla 164).

Los bosques estudiados en el sector Sur fueron descritos en su composición florística y estructura recientemente por Rangel & Avella (2011) y Estupiñán *et al.* (2011) y agrupados en cuatro alianzas y ocho asociaciones, y los relictos de bosque estudiados en el sector Noroccidente hacen parte de cuatro asociaciones descritas por Rangel *et al.* (2010), que se enuncian en la Tabla 164. Los sitios de muestreo se seleccionaron de tal manera que incluyeran variación en las condiciones climáticas, principalmente en la precipitación la cual varía entre 1340 y 2900 mm anuales, es decir se estudiaron bosques en un gradiente que va desde clima súper húmedo a semi-húmedo. Además incluyen variación en las características topográficas (Tabla 164).

**Toma de información**

En cada localidad se establecieron parcelas de 500 m$^2$ en las cuales se evaluaron aspectos de la composición florística y estructura de la vegetación siguiendo la metodología descrita por Rangel & Velásquez (1997).

En cada levantamiento se midieron todas las plantas leñosas (árboles y palmas), con DAP ≥ 5 cm. Los parámetros registrados fueron nombre de la especie, el DAP (diámetro a la altura del pecho o a 1,3 m sobre el suelo o encima de las bambas en caso de que las tuviera), altura total y altura del tallo y cobertura. El DAP se midió con cinta diamétrica, la altura se determinó por estimación visual con ayuda de una vara graduada y la cobertura se estimó calculando directamente el área que proyecta sobre el suelo la copa de cada individuo. De cada especie se colectó material vegetal (fértil y estéril) usando bajarramas, el cual se preparó, se montó en papel periódico y se preservó usando alcohol etílico al 75%. Se registró información característica de los individuos (nombre común, uso, hábitat, forma de crecimiento, tamaño, color de las flores y/ frutos, aspectos de la corteza como color, olor, exudado, rasgos de las hojas como forma, protección, entre otros). El material recolectado y la información registrada se utilizaron para la determinación de las especies en campo y/o en el herbario Nacional Colombiano del Instituto de Ciencias Naturales.

El estudio se realizó con base en mediciones en campo y se complementó con información primaria de la caracterización realizada por Rangel *et al.* (2010) para los bosques circundantes a los complejos de humedales del departamento de Córdoba, la caracterización florística y estructural de los bosques del Sur del departamento de Córdoba (Rangel & Avella, 2011) y el estudio sobre la estructura y la composición florística de los bosques inundables del parque nacional natural Paramillo (Estupiñán *et al.*, 2011) como se enuncia en la Tabla 164.

**• Peso específico de la madera**

Para medir esta variable se seleccionaron cuatro tipos de bosque de acuerdo con el patrón de lluvias descrito por Rangel & Carvajal (2012). Bosques de la asociación Jacarando copaiae - Pouterietum multiflorae, lozalizados en áreas con clima súper húmedo (precipitación entre >2200-3000 mm anuales); bosques de la asociación Protio aracouchini - Viroletum elongatae ubicados en clima muy húmedo (precipitación entre >1800-2200 mm/anuales); bosques de la asociación Trichilio hirtae - Schizolobietum parahibi en clima húmedo (>1400-1800 mm anuales) y bosques de la comunidad de *Acalypha* sp. y *Guazuma ulmifolia* ubicados en clima semihúmedo (>1000-1400 mm/anuales).

Para la determinación del peso específico se tomaron tres (3) muestras de madera del tallo de 5 cm$^3$ de cada individuo con diámetro mayor a 5 cm presente en los levantamientos. Las muestras se tomaron a 1.3 m del suelo y para extraerlas se utilizó un taladro con copa de 1 pulgada, manteniendo en lo posible la misma profundidad. Se pesaron en el campo usando una balanza con precisión de 0.01 g, se envolvieron en papel periódico, se etiquetaron, se preservaron en alcohol etílico al 75% y se almacenaron en bolsas plásticas para su transporte al laboratorio. Allí se determinó el volumen de cada muestra mediante medición indirecta, para lo cual se sumergieron en una probeta graduada que contenía agua con un volumen conocido y se registró el volumen de agua desplazada, el cual equivale al volumen de la muestra. Posteriormente, las muestras se secaron en el horno del laboratorio de Palinología y Paleoecología de la Universidad Nacional de Colombia a 105°C hasta peso constante para obtener la masa anhidra (seca) y se pesaron en una balanza electrónica de precisión 0,01 g.

**• Contenido de carbono**

El contenido de carbono se midió en los individuos leñosos presentes en los levantamientos seleccionados (DAP> 5cm); a cada uno se le tomaron tres muestras de cada componente (tallo, rama, corteza y hojas). Las muestras de tallos que se tomaron para determinar el peso específico de la madera se utilizaron también para evaluar el contenido de carbono en este componente. Las muestras de corteza se tomaron con machete y las de ramas y hojas se colectaron con bajaramas. Se siguió el mismo procedimiento de preservación y secado, que se realizó con las muestras para la determinación del peso específico.

**Tabla 164.** Información general de las localidades de estudio y de los levantamientos de vegetación realizados en el gradiente de precipitación: súper húmedo-semi húmedo y tipos de vegetación correspondiente.
Lev: levantamiento; PMA: precipitación media anual.

| SÍMBOLO | FORMACIÓN VEGETAL DEL SITIO | ALIANZA | ASOCIACIÓN | LEV | LOCALIDAD | LATITUD | LONGITUD | ALTITUD (m) | PENDIENTE | PMA (mm) | AUTOR-FUENTE |
|---|---|---|---|---|---|---|---|---|---|---|---|
| BHtf2/Jco-Pmu | Bosque tropical primario en clima súper húmedo de poco a medianamente intervenido | Brosimo utilis - Pentaclethrion macrolobae | Jacarando copaiae-Pouterietum multiflorae | SC-1 | Montelibano, Candelaria, Lagundera | 7.799139 | -75.8552 | 136-334 | 7.62 | 2422 | Avella & Rangel, 2011 |
| | | | | SC-10 | Tierralta, La Oscurana, Nueva Estrella | 8.008556 | -76.0899 | | 16.95 | 2102 | |
| | | | | SC-2 | Montelibano, Aguasprietas, Guájaro | 7.769222 | -75.876 | | 5.91 | 2509 | |
| Bmhtf2/Mla-Pma | Bosque tropical primario en clima muy húmedo medianamente intervenido | | Marilo laxiflorae-Pentaclethretum macrolobae | SC-6 | Tierralta, La Oscurana, Onomás | 8.012361 | -76.1028 | 157-174 | 15.8 | 2097 | |
| | | | | SC-9 | | 8.007611 | -76.1015 | | 11.62 | 2094 | |
| Bmhtf1/Par-Vel | Bosque tropical primario en clima muy húmedo poco o nada intervenido | | Protio aracouchini-Viroletum elongatae | SC-16A | Tierralta, Tuis-tuis, Tuti-fruti | 8.036944 | -76.0964 | 136-178 | 8.25 | 2021 | Este estudio |
| | | | | SC-16B | | 8.036944 | -76.0956 | | 21.06 | 2021 | |
| | | | | SC-16C | | 8.037858 | -76.096 | | 12 | 2021 | |
| BHtf1/Twe-Qhu | Bosque tropical primario en clima super húmedo poco o nada intervenidos | Billio roseae-Quercion humboldtii | Tovomito weddellianae-Quercetum humboldtii | SC-17 | Tierralta, El silencio - PNN Paramillo | 7.960194 | -76.0842 | 774-937 | 12.34 | 2309 | Rangel & Avella (2011). |
| | | | | SC-18 | | 7.960278 | -76.0906 | | 13.43 | 2422 | |
| | | | | SC-20 | | 7.960944 | -76.0839 | | 14.04 | 2309 | |
| BHin1/Pde-Tpo | Bosque tropical primario en clima super húmedo en sitios inundables poco o nada intervenidos | Eschweilero coriaceae - Pentacletrion macrolobae | Prestoeo decurrentis-Trichillietum poeppigi | PNP-1 | Tierralta, Llanos Río Tigre - PNN Paramillo | 7.6115 | -76.0142 | 179-213 | 1.13 | 2899 | Estupiñan et al., 2011 |
| | | | | PNP-2 | | 7.646306 | -76.0111 | | 0 | 2842 | |
| | | | | PNP-3 | | 7.619111 | -76.0121 | | 1.52 | 2899 | |
| | | | | PNP-4 | | 7.658278 | -76.0243 | | 3.24 | 2808 | |
| BHri1/Mis-Ppu | Bosque tropical primario en clima super húmedo en sitios asociados a cuerpos de agua poco o nada intervenidos | | Macrolobio ischnocalycis-Peltogynetum purpurea | PNP-6 | Tierralta, Río Manso - PNN Paramillo | 7.662167 | -76.0962 | 196-246 | 4.08 | 2868 | |
| | | | | PNP-7 | | 7.666111 | -76.0945 | | 4.52 | 2849 | |
| | | | | PNP-8 | | 7.673917 | -76.1109 | | 4.08 | 2798 | |
| | | | | PNP-9 | | 7.665861 | -76.1021 | | 0.72 | 2868 | |
| Bmhtf2/Cpy-Pdo | Bosque tropical primario en clima muy húmedo poco intervenido | Astrocaryo malybo - Cavanillesion platanifoliae | Cariniano pyriformis-Pentaplarietum doroteae | SC-12 | Valencia, Nuevo Oriente, B. Comunitario | 8.1625 | -76.2083 | 127-124 | 6.08 | 1818 | Avella & Rangel, 2011 |
| | | | | SC-13 | | 8.162861 | -76.2074 | | 4.66 | 1818 | |
| Bmhtf2/Mgr-Ama | Bosque tropical primario y secundario en clima muy húmedo muy intervenido | | Mayno grandifoliae-Astrocaryetum malybo | SC-11 | | 8.162056 | -76.2078 | 97-142 | 4.52 | 1818 | |
| | | | | SC-14 | | 8.148194 | -76.2333 | | 0.51 | 1860 | |
| Bhtf2/Cpr-Cpa | Bosque tropical primario en clima húmedo de poco a medianamente intervenido | Cratevo tapiae - Cavanillesion platanifoliae | Cordietum proctato-panamensis | NO-P25 | Los Córdobas, Floral, La Ceba | 8.769028 | -76.3046 | 97-139 | 2.53 | 1441 | Rangel et al., 2010 |
| | | | | NO-P23 | Canalete, Cordobita Central, El Chimborazo | 8.746417 | -76.3198 | | 2.53 | 1454 | |
| | | | | NO-P30 | | 8.751333 | -76.3302 | | 2.86 | 1477 | |
| | | | | NO-P31 | | 8.751361 | -76.3302 | | 1.6 | 1477 | |
| | | | | NO-P32 | | 8.751389 | -76.3303 | | 1.6 | 1477 | |
| | | | | NO-P33 | | 8.749389 | -76.317 | | 3.24 | 1454 | |
| | | | | NO-P34 | | 8.749333 | -76.3171 | | 3.2 | 1454 | |
| | | | | NO-P35 | | 8.749222 | -76.317 | | 5.27 | 1454 | |
| | | | | NO-P36 | | 8.768556 | -76.3118 | | 5.08 | 1441 | |
| | | | | NO-P37 | | 8.770583 | -76.3097 | | 1.13 | 1441 | |
| | | | | NO-P38 | | 8.766722 | -76.309 | | 2.02 | 1441 | |
| Bhtf2/Cod-Cpl | Bosque tropical primario en clima húmedo medianamente intervenido | | Cappari odoratissimatis-Cavanillesietum platanifoliae | NO-P20 | | 8.745944 | -76.3269 | 98-111 | 2.86 | 1468 | |
| | | | | NO-P21 | | 8.740167 | -76.3399 | | 2.09 | 1487 | |
| | | | | NO-P22 | | 8.743917 | -76.324 | | 5.44 | 1468 | |
| Bhtf1/Thi-Spa | Bosque tropical en clima húmedo primario poco intervenido | Alianza no definida | Trichilio hirtae-Schizolobietum parahibi | NC-3 | Los Córdobas, S. Rosa Caña, Campo Alegre | 8.795139 | -76.3235 | 100-177 | 5.15 | 1460 | Este estudio |
| | | | | NC-4 | | 8.794278 | -76.3204 | | 13.06 | 1444 | |
| | Bosque tropical en clima húmedo primario medianamente intervenido | | | NO-P1 | Los Córdobas, S. Rosa Caña, Nuevo Nariño | 8.7995 | -76.3303 | | 5.27 | 1478 | Rangel et al., 2010 |
| | | | | NO-P4 | | 8.797667 | -76.3239 | | 10.62 | 1460 | |
| | | | | NO-P5 | | 8.798639 | -76.3249 | | 6.57 | 1465 | |
| | | | | NO-P6 | | 8.795667 | -76.3318 | | 7.21 | 1482 | |
| Bttf3/Asp-Gul | Bosque tropical en clima semi húmedo muy intervenido | Alianza no definida | Acalypha sp. y Guazuma ulmifolia | NC-1 | Moñitos, La Vega, Viento Solar | 9.121444 | -76.1859 | 10-147 | 3.2 | 1400 | Este estudio |
| | | | | NO-P24 | Pto. Escondido, Santa Bárbara | 8.737806 | -76.2232 | | 1.52 | 1343 | Rangel et al., 2010 |
| | | | | NO-P26 | Canalete, La Lorenza, La Pozada | 9.097361 | -76.2055 | | 3.2 | 1403 | |

En el Laboratorio de carbones del Instituto de Investigaciones en Geociencias y Minería INGEOMINAS, las muestras secas se molieron dos veces hasta un obtener un tamaño de 0.25 mm, luego se secaron nuevamente a 105°C durante 24 horas en el horno del laboratorio de Palinología de la Universidad Nacional de Colombia, sede Bogotá, para eliminar cualquier resto de humedad. Posteriormente, una pequeña cantidad (0.020 g) de cada muestra se colocó en la cámara del analizador elemental de carbono, hidrógeno y nitrógeno Leco modelo TruSpec (Leco Corporation 1984), el cual mediante el método de combustión en seco a 950°C permite obtener el porcentaje de carbono por unidad de masa seca presente en cada muestra. Este servicio fue prestado por el Laboratorio de Carbones del Instituto de Investigaciones en Geociencias y Minería INGEOMINAS.

**Procesamiento de la información**

**Estructura**

Se realizó una clasificación de levantamientos y especies a partir del índice de predominio fisionómico, mediante Twinspan, procedimiento que permitió ubicar los levantamientos en cada comunidad de vegetación como se observa en la Tabla 164.

Los árboles se clasificaron en seis clases de acuerdo con su DAP: I: 2.5-10cm, II: >10-30cm, III: >30-50cm, IV: >50-70, V: >70-100 y VI:>100cm. Se utilizó esta clasificación con fines comparativos con otros estudios. También se utilizó la clasificación sugerida por Rangel & Velásquez (1997) para estudiar la distribución vertical de la vegetación, con los siguientes intervalos de altura: arbustivo (ar): 1.5-5m; subarbóreo (Ar): 5-12m; arbóreo inferior (Ai): 12-15 y arbóreo superior (As): >25m.

Se realizaron distribuciones de frecuencia de la biomasa por clase diamétrica con intervalos de 10 cm y por estrato para cada tipo de bosque y se utilizó la regla de Sturges para obtener clases de variación de la biomasa y carbono almacenados en la vegetación.

**Estimación de la biomasa**

Para estimar la biomasa aérea de los bosques se usan generalmente dos métodos: métodos directos destructivos y métodos indirectos que realizan simulaciones a partir de información básica de inventarios y de imágenes satelitales. Los métodos destructivos son básicamente: 1. Cosecha de la totalidad de la vegetación, 2. Estimación de la biomasa aérea por el método del árbol medio y 3. Modelos de regresión, que relacionan la masa seca de algunos árboles con otras variables (DAP, altura, peso específico de la madera, entre otras). Los modelos obtenidos se utilizan para estimar la biomasa del árbol, en un área conocida. El método de los modelos de regresión es la mejor aproximación, y por ende se usa en la mayoría de investigaciones de cuantificación de biomasa de los bosques tropicales (Zapata *et al.*, 2003). Para este fin, uno de los modelos más utilizados es el modelo alométrico, el cual relaciona la masa seca del árbol con una variable de fácil medición.

En el presente estudio, para la estimación de biomasa aérea, se usó el método de modelos alométricos derivados de una muestra de árboles cosechados en áreas geográficas con condiciones similares a las del estudio, como lo recomiendan varios autores.

**• Parámetros usados**

Para la estimación de la biomasa aérea de los individuos con diámetro mayor a 5 cm se utilizaron los datos de parámetros estructurales medidos en los levantamientos (DAP y altura). Además, se incluyó el peso específico aparente básico de la madera ρ, definido como el cociente entre el peso de la madera anhidra (aquella en la que se ha eliminado toda la humedad) y el volumen en condiciones de campo (estado verde). A partir de estos valores se calculó ρ con la siguiente fórmula:

$\rho = Pa/Vv$

Siendo ρ= Peso específico aparente básico (g/cm3)
Pa= peso de la madera en estado anhidro (g)
Vv: volumen de la madera en estado verde (cm3)

Debido a las dificultades logísticas, no fue posible medir todas las especies presentes en los levantamientos de vegetación, por lo tanto, el peso específico de la madera se midió solamente en los bosques seleccionados; la información de las otras especies presentes en los demás bosques se obtuvo de valores registrados en la literatura (Global Wood Density Database, Chave *et al.*, 2009; Lastra, 1975 y Anzola & Rodríguez, 2001). Para las especies que no se encontraron registros en las bases de datos consultadas, o aquellas cuya determinación fue solo hasta el nivel de género, se usó el promedio de valores de las especies medido y/o mencionado, pertenecientes a dicho género. Cuando no se obtuvo información a nivel de género se usó el promedio de valores medidos para cada localidad.

*Ecuaciones alométricas*

La elección del modelo para estimar la biomasa es crucial, para evitar errores en la estimación. Varios autores sugieren utilizar ecuaciones generadas localmente, sin embargo cuando no se dispone de estas, se pueden emplear ecuaciones ya existentes, generadas en condiciones climáticas y edáficas similares a la de la zona de estudio, especialmente cuando se han validado con datos de cosecha. En el presente estudio se estimó la biomasa aérea de árboles ≥ 5cm de DAP mediante nueve modelos alométricos mencionados en la literatura, que se usan comúnmente en investigaciones con condiciones similares a la zona de estudio (modelos 1-9 en la tabla 165). Estos modelos estiman la biomasa en función de variables de fácil medición tales como diámetro (DAP), altura del árbol (H) y peso específico ($\rho i$) de la madera. De acuerdo con las variables predictivas usadas para la estimación, se usaron cuatro series de ecuaciones a saber: (a) en función del DAP (modelos 1-2), (b) en función del DAP y altura, (3 y 4), (c) en función del DAP y peso específico de la madera (5 y 6), y (d) en función del DAP, altura y peso específico de la madera (modelos 7, 8 y 9).

Para la estimación de la biomasa de los árboles de gran porte (individuos con DAP> 156 cm) los cuales superan el límite de diámetro que acepta el modelo seleccionado, se generó una ecuación polinomial de grado diez (Tabla 165) a partir de los datos DAP, altura y peso específico de la madera y biomasa estimada usando el modelo de Chave *et al.* (2005). Esta ecuación permite obtener valores menores de biomasa para los árboles grandes si se le compara con los que se obtienen con el modelo de Chave *et al.* (2005), debido a que se obtuvo a partir de los datos locales.

La biomasa de los individuos con DAP menor a 5 cm, se calculó usando la ecuación propuesta por Hughes (1999) y la corrección propuesta por Chave (2003), que incorpora el peso específico de la madera. Se utilizó para esta categoría un promedio de valores de peso específico de la madera que proviene de los individuos medidos entre 3 y 7 cm de diámetro, específico para cada localidad de muestreo.

Una situación bastante frecuente cuando se estima la biomasa aérea en los bosques tropicales es que los modelos tienden a sobreestimar los árboles de gran porte, como se observa en la figura 375 con los individuos con DAP >80 cm. Esta situación, sumada al tamaño de las parcelas genera inconvenientes al convertir la biomasa calculada en las parcelas de muestreo (500 m$^2$) a hectáreas, este proceder conduce a que se propague el error y se sobreestima la biomasa. Además el número de individuos de las clases diamétricas mayores tiende a ser muy bajo y al escalar a hectáreas puede sobreestimarse también la cantidad de estos individuos y por ende la biomasa. Por estas razones, en el presente estudio no se realizó la conversión de la unidad de muestreo (parcelas de 500 m$^2$) a hectáreas sino que los resultados se presentan para el tamaño de área muestreada (0.05 ha). Esta información se está complementando con datos de densidad por hectárea para generar índices para la zona que permitan realizar las estimaciones en áreas mayores con los datos ajustados a las condiciones locales, y de esta manera realizar los cálculos a nivel regional de una forma precisa; dicho procedimiento hace parte de otro estudio (Biól. H. Arellano, com. pers., I.C.N.).

**Cálculo del carbono almacenado en la biomasa**

Para determinar el carbono almacenado en la biomasa es habitual multiplicar la biomasa y la fracción de carbono contenida en ella; no obstante en el presente estudio se generó información de carbono para los individuos presentes en algunas localidades de muestreo y para cuatro tejidos del árbol dada la variación que se presenta entre estos; entonces fue necesario determinar la distribución de la biomasa en los diferentes componentes del árbol y así relacionar la información de masa y carbono. Se utilizaron las ecuaciones de Overman *et al.* (1990) para estimar la biomasa por componente del árbol (Tabla 166) y usar esta proporción para el cálculo de carbono en la biomasa para cada componente.

Se multiplicó la fracción de carbono medida en cada tejido (tallo, ramas, corteza y hojas) y la cantidad de biomasa de cada componente del árbol para obtener la reserva de carbono por componente; la suma de todos los componentes da como resultado la reserva de carbono en todo el individuo. Los valores de todos los individuos presentes en cada levantamiento se sumaron para obtener la reserva de carbono en la biomasa de todo el levantamiento.

**Tabla 165.** Modelos alométricos empleados para estimar la biomasa aérea de árboles con DAP≥2.5cm de los bosques del Sur y Noroccidente de Córdoba, Colombia.

| Autor | Modelo | Procedencia | Amplitud DAP |
|---|---|---|---|
| 1. IPCC (2005) | $B = \exp[-2.289+2.649*\ln(DAP)-0.021*\ln(DAP^2)]$ | | |
| 2. Chave *et al.* (2001) | $\ln B = [-2.19+2.54*\ln(DAP)]$ | Bosque húmedos tropicales | >10 cm |
| 3. Rodríguez (1989) | a) Peso seco madera tallo $=7.54908+0.01753*DAP^2*Hc +0.10827*10-19*DAP^2*H$. <br> b) Peso seco corteza tallo $=0.98155+0.00229*(DAP^2*H)$. <br> c) Peso seco madera rama $= 18.54992-(0.29015*10-3)*(DAP^2*Hc)$ <br> d) Peso seco corteza rama $=5.09423+0.27629*10-7*(DAP2*H)^2$ <br> e) Peso seco madera ramilla $= -2.09893+0.5409*DAP-0.0357*H$ <br> f) Peso seco corteza ramilla$=-0.375+0.0012*(DAP^2*H)$ <br> g) Peso seco hojas $=0.53331+0.00103*(DAP^2*H)$ <br> $BA = a + b + c + d + e + f + g$ | Bosque pluvial tropical, bajo Calima, Buenaventura, Colombia. | ≥3 cm |
| 4. Overman *et al.* (1990) | $\ln = -3.84 +1.035*\ln(DAP^2*H)$ | Bosque pluvial tropical, Caquetá, Colombia | 8.1-100 cm |
| 5. Overman *et al.* (1990) | $\ln B = [-0.906+1.77*\ln((DAP)^2* \rho i)]$ | | 8.1-100 cm |
| 6. Chave *et al.* (2005) | $B = \rho i * \exp(-1.499+2.148*\ln(DAP)+ 0.207 (\ln(DAP))^2 -0.0281(\ln(DAP))^3)$ <br> $B = \rho i * \exp(-0.667 + 1.784*\ln(DAP) + 0.207(\ln(DAP))^2 - 0.0281(\ln(DAP^3)))$ | Bosques húmedos tropicales mundiales <br> Bosques secos tropicales mundiales | 5-156 cm <br> 5-156 cm |
| 7. Anzola y Rodríguez (2001) | $B = VT * \rho i * FEB$ ajustado | Bosques de Colombia | 4-112 cm |
| 8. Overman *et al.* (1990) | $\ln B = [-2.904 +0.993*\ln((\rho i (DAP^2)H)]$ | Bosque pluvial tropical, Caquetá, Colombia | 8.1-100 cm |
| 9. Chave *et al.* (2005) | $B = \exp[-2.977 + \ln(\rho i (DAP)^2 H)]$ <br> $B = \exp[-2.187 + 0.916 * \ln(\rho i (DAP)^2 H)]$ | Bosque húmedos tropicales mundiales <br> Bosques secos tropicales mundiales | 5-156 cm <br> 5-156 cm |
| Este estudio | $B = -55.39+5.831(DAP)+24.64(H)-1.658(DAP*H)-4.776*(H^2)+0.2534(DAP*H^2)+0.4202(H^3)-0.007398(DAP*H^3)-0.02141(H^4)+0.0001497(DAP*H^4)+0.0002943(H^5)$ | Bosques húmedos y semihúmedos de Córdoba, Colombia | >156 cm |
| 10. Hughes (1999) | $B = [-1.839 +2.116 \ln(DAP)] * \rho i$ | Bosques secundarios, México | 1-10 cm |

Convenciones: B: biomasa; DAP: diámetro a la altura del pecho (1.30 m); H: altura total; Hc: altura comercial; $\rho_i$: peso específico de la madera (g/cm$^3$), i=valor específico para cada especie, o promedio local cuando no se conoce el valor para la especie. VT: volumen del tallo (m$^3$/ha); FEB: factor de expansión de biomasa; ln: logaritmo neperiano.

**Tabla 166.** Modelos para la estimación de biomasa por componente del árbol (Overman *et al.*, 1990).

| Componente | Ecuación |
|---|---|
| BA tronco | lnB= -2.881+0.956 Ln ((DAP)$^2$)hρ) |
| BA ramas | lnB= -7.765+1.304 Ln ((DAP)$^2$)hρ) |
| BA ramitas | lnB= -2.612+0.659 Ln ((DAP)$^2$)hρ) |
| BA hojas | lnB= -2.444+0.478 Ln ((DAP)$^2$)hρ) |

Convenciones: B: biomasa; DAP: diámetro a la altura del pecho (1,30 m); H: altura total; ρ: peso específico de la madera (g/cm3); ln: logaritmo neperiano.

**Análisis de los datos**

Para evaluar la estructura de la vegetación y la variación de la biomasa en los bosques del Sur y Noroccidente de Córdoba, se realizaron distribuciones de frecuencia y se calcularon los estadísticos descriptivos (medidas de tendencia central y de dispersión) de las variables estructurales: densidad (número de individuos), área basal, altura y de la biomasa en cada tipo de bosque, a nivel general y por clase diamétrica. Se probaron los supuestos de normalidad mediante las pruebas de Kolmogorov-Smirnov y Shapiro-Wilk cuyos estadísticos son D y W respectivamente, que se basan en la máxima diferencia entre la distribución acumulada de la muestra y la distribución acumulada hipotética; se contrasta la hipótesis nula Ho: la distribución de los datos de la muestra se ajustan a una distribución normal; si los estadísticos D y W, respectivamente son significativos, entonces la hipótesis Ho debe rechazarse.

Se utilizó el coeficiente de correlación de Spearman (R) para evaluar la predicción que cada modelo hace de los resultados de los otros modelos empleados para la estimación de la biomasa, siguiendo la propuesta de Chave *et al.* (2003) y de esta manera seleccionar el modelo que presente la mayor correlación media con los otros modelos que se están comparando. El coeficiente de correlación de Spearman es una medida de asociación entre dos variables aleatorias continuas. En su cálculo los datos se ordenan y se reemplazan por su respectivo orden, mediante la siguiente expresión:

$$r_{S_{(x,y)}} = 1 - 6 \frac{\sum D^2}{n(n^2-1)}$$

Donde $D$ es la diferencia entre los estadísticos de orden de las dos variables a comparar (x-y), n es el número de parejas.

Se realizó un análisis de varianza de una vía no paramétrico (prueba de Kruskal - Wallis) y comparaciones múltiples, con el fin de evaluar las diferencias significativas de las variables estructurales (área basal, altura, peso específico de la madera y número de individuos) y la biomasa total por asociación y por clase diamétrica entre los diferentes tipos de bosque. La prueba de Kruskal-Wallis también conocida como ANOVA por categorías se utilizó para contrastar la hipótesis Ho: las *j* muestras cuantitativas (categorías) de cada tipo de bosque se obtuvieron de la misma población o de poblaciones con medianas iguales; la prueba para este fin asigna categorías a las observaciones, suma estas categorías separadamente para cada grupo y las compara calculando el estadístico H como se enuncia a continuación:

$$H = \frac{12}{n(n=1)} \sum_{j=1}^{J} \frac{R_j^2}{n_j} - 3(n+1)$$

Donde H = valor estadístico de la prueba de Kruskal-Wallis, N = tamaño total de la muestra, $R_j^2$ = sumatoria de los rangos elevados al cuadrado, $n_j$ = tamaño de la muestra de cada grupo, J = muestras aleatorias. El estadístico H calculado se compara con valores críticos de la distribución ji cuadrado con J-1 grados de libertad. Si la hipótesis nula es cierta y las categorías se han extraído de la misma población (o de poblaciones con medianas idénticas) y todas las muestras tienen puntuaciones de 5 o más, entonces la distribución muestral de H es una aproximación muy cercana de la distribución ji-cuadrado. Si las poblaciones no son diferentes una de la otra, se espera que las sumas de sus categorías sean aproximadamente las mismas. Una vez que se ha rechazado la hipótesis nula con la prueba de Kruskal-Wallis, y se admite que al menos una de las *j* muestras no pertenece a la misma población, se está interesado en saber cuál de las *j* muestras tiene una mediana diferente de las demás; para este fin se realizaron las comparaciones múltiples en las cuales se contrastan los rangos medios de todos los pares de grupos mediante el cálculo del estadístico *z'* para cada comparación entre los grupos (tipos de bosque); se obtiene un valor p asociado con cada comparación, p es la probabilidad asociada con *z'* (Siegel & Castellan, 1988).

Para el procesamiento general de los datos y la estimación de biomasa y carbono se utilizó el programa Excel; para el análisis Twinspan se usó el programa PcOrd4 (McCune & Mefford, 2006) y para las demás pruebas estadísticas y gráficas se empleó el paquete estadístico STATISTICA (StatSoft Inc, 2002).

**RESULTADOS**

**Estimación de la biomasa**

Las estimaciones realizadas según los diferentes modelos presentaron diferencias que se corroboraron mediante la prueba de Kruskal-Wallis (H (8, N= 180) =34.53 p =0.00).

En la figura 375 se muestra la distribución de la biomasa en cada clase diamétrica con cada uno de los nueve modelos empleados para su estimación. Los modelos 1, 2, 5 y 6 presentan las estimaciones de biomasa más altas en todas las clases diamétricas. Los modelos 1 y 2 estiman la biomasa en función del DAP y los modelos 5 y 6 estiman la biomasa en función del DAP y del peso específico de la madera ($\rho$). El modelo 4 que incorpora en sus variables predictivas al DAP y la altura, presentó valores medios de biomasa. Los modelos 7, 8 y 9 involucran como variables predictivas al DAP, la altura y el peso específico de la madera, presentan estimaciones medias que son bastante similares. La menor estimación se obtuvo con el modelo 3, el cual estima la biomasa en función del DAP y altura. Las diferencias entre los modelos pueden deberse a que árboles con el mismo diámetro presentan valores diferentes de altura y/o peso específico de la madera.

Las correlaciones entre los valores de biomasa estimados por los diferentes modelos fueron significativas ($r^2$>0.80 a excepción del modelo 3 con $r^2$=0.58). Las estimaciones de biomasa realizadas con los modelos 1, 2, 8 y 9 presentaron las mayores correlaciones en comparación con los modelos restantes (Tabla 167).

Al observar las fórmulas alométricas evaluadas y las altas correlaciones (tabla 167) de sus resultados, se llega a la conclusión que las más actuales usan la misma estructura matemática, en otras palabras, se incluyen pequeñas modificaciones a un modelo básico. Al comparar los resultados de las estimaciones para todos los modelos entre elementos pequeños y grandes, se observa que las zonas que representan los resultados para las clases inferiores acumulan menos error que los registros estimados en clases superiores (Figura 376) y presentan tendencias similares. Esta razón impulsa el fortalecimiento de estimaciones que evalúen el problema desde diferentes perspectivas, se requieren otros modelos que permitan minimizar el error existente y así calibrar los resultados obtenidos con las metodologías tradicionales con el fin de obtener registros más precisos.

La mayoría de los modelos utilizan el DAP como variable estimadora de la biomasa; sin embargo, en la literatura se ha resaltado la importancia de incorporar también en los modelos, la altura y el peso específico de la madera, dado que éstos parámetros son responsables de un alto porcentaje de la variación de la biomasa en los bosques tropicales (Chave *et al.*, 2005). Por lo tanto, entre los cuatro modelos que presentaron las mayores correlaciones (modelos 1, 2, 8 y 9), se seleccionaron los modelos 8 y 9 que integran las variables DAP, altura y peso específico de la madera; con la expectativa de que permitieran estimar con mayor precisión la biomasa. Los modelos 8 y 9 presentaron estimaciones muy similares, por consiguiente se seleccionó el 9 bajo el criterio de que fue generado a partir de un mayor número de muestras, es aplicable para una categoría de árboles amplia (5-156cm de DAP) y además, discrimina entre bosques húmedos y secos. Con este modelo se realizaron las estimaciones de todos los árboles con diámetro entre 5 y 156 cm en su DAP, en los diferentes tipos de bosque a lo largo del gradiente de precipitación de Sur a Norte del departamento de Córdoba.

**Estructura, biomasa y carbono por tipo de bosque**

**Estructura y biomasa**

La variación de la biomasa y el carbono con respecto a la estructura en los diferentes tipos de bosque, presentó las siguientes particularidades:

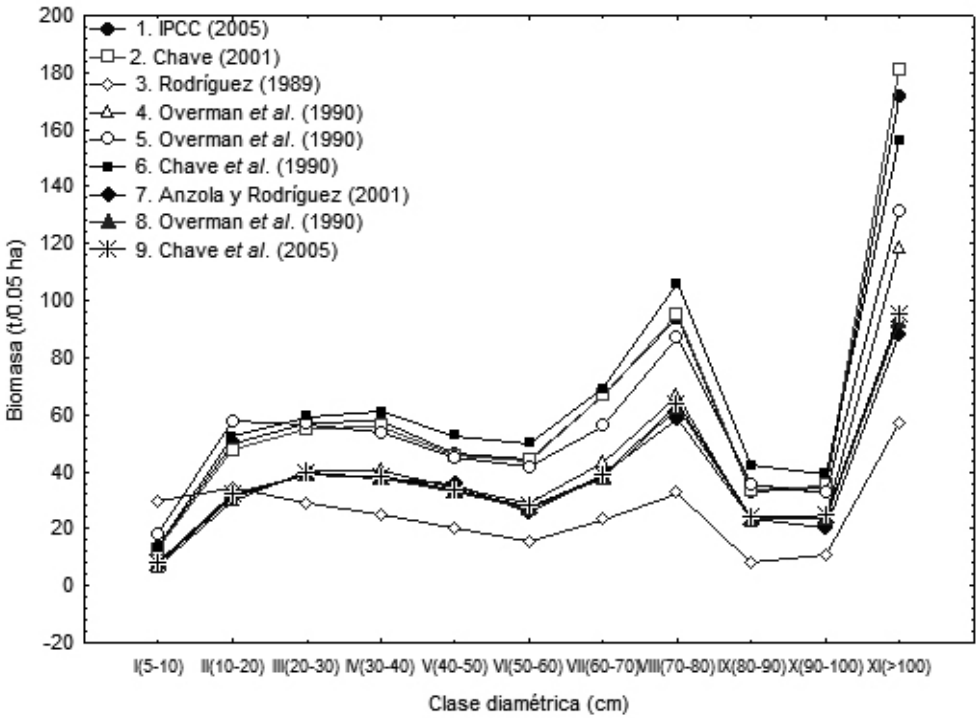

**Figura 375.** Distribución de biomasa aérea por clase diamétrica (cm) usando nueve modelos alométricos mencionados en la literatura.

Convenciones:

| Código modelo | Modelo |
|---|---|
| 1. IPCC (2005) | $B= \exp[-2.289+2.649*\ln(DAP)-0.021*\ln(DAP^2)]$ |
| 2. Chave et al. (2001) | $\ln B=[-2.19+2.54*\ln(DAP)]$ |
| 3. Rodríguez (1989) | $B=B_{tallo}+B_{ramas}+B_{corteza}+B_{hojas}$ |
| 4. Overman et al. (1990) | $\ln= -3.84 +1.035*\ln(DAP^2*H)$ |
| 5. Overman et al. (1990) | $\ln B= [-0.906+1.77*\ln((DAP)^2* \rho i)]$ |
| 6. Chave et al. (2005) | $B= \rho i * \exp(-0.667 + 1.784*\ln(DAP) + 0.207(\ln(DAP))^2 - 0.0281(\ln(DAP))^3)$ |
| 7. Anzola y Rodríguez (2001) | $B= VT * \rho i * FEB$ ajustado |
| 8. Overman et al. (1990) | $\ln B= [-2.904 +0.993*\ln((\rho i (DAP^2)H)]$ |
| 9. Chave et al. (2005) | $B= \exp[-2.187 + 0.916 * \ln(\rho i(DAP)^2 H)]$ |

La biomasa estimada para los doce tipos de bosque varió entre 7.39±1.79 t/0.05 ha para los levantamientos de la comunidad de *Acalypha* sp. y *Guazuma ulmifolia* y 41.15±21.81 t/0.05ha para la asociación Macrolobio ischnocalycis - Peltogynetum purpurea. El carbono almacenado en la biomasa fluctúa entre 3.34 y 18.72 t/0.05 ha para estos tipos de bosque. Estas estimaciones tuvieron una incertidumbre de estimación asociada del 17.41%.

A partir de los resultados obtenidos en las estimaciones de biomasa y carbono, se organizaron mediante la regla de Sturges las siguientes clases de variación de los datos con respecto a las reservas de biomasa y carbono: *muy altas, altas, medias, bajas y muy bajas* (Tabla 168). En la clase *muy altas* (>36.79-44.14) se ubican las asociaciones Prestoeo decurrentis - Trichillietum poeppigi, Macrolobio ischnocalycis - Peltogynetum purpurea con valores de biomasa que oscilan entre 40.83 y 41.15 t/0.05 ha y de carbono entre 18.44 y 18.9 t/0.05ha. Los bosques de las asociaciones Protio aracouchini - Viroletum elongatae y Tovomito weddellianae - Quercetum humboldtii pertenecen a la clase *altas* (>29.44-36.79), dado que presentan valores de biomasa entre 30.01 y 34.33 t/0.05ha y carbono entre 13.78 y 15.76 t/0.05ha. Los bosques de las asociaciones Trichilio hirtae-Schizolobietum parahibi y Cappari odoratissimatis - Cavanillesietum platanifoliae presentan valores de biomasa de 24.06 y 22.45 y t/0.05ha, respectivamente y corresponden a la clase reservas de biomasa *medias* (>22.09-29.44). En la categoría *bajas* (>14.74-22.09) se ubicaron las asociaciones Cariniano pyriformis - Pentaplarietum doroteae y Jacarando copaiae - Pouterietum multiflorae y Jacarando copaiae - Pouterietum multi-

**Tabla 167.** Matriz de correlación de Spearman de los nueve modelos utilizados para la estimación de biomasa aérea.

Los valores corresponden al coeficiente de correlación. Se emplearon las estimaciones de biomasa aérea realizadas en los bosques del Sur de Córdoba, para hacer las correlaciones. Los modelos y su referencia se enuncian en la tabla 168. La última fila corresponde al valor medio de correlación de cada modelo.

| | 1. IPCC (2005) | 2. Chave (2001) | 3. Rodríguez (1989) | 4. Overman (1990) | 5. Overman *et al.* (1990) | 6. Chave (2005) | 7. Anzola y Rodríguez (2001) | 8. Overman *et al.* (1990) | 9. Chave *et al.* (2005) |
|---|---|---|---|---|---|---|---|---|---|
| 1. IPCC (2005) | 1.000 | 1.000 | 0.591 | 0.991 | 0.909 | 0.982 | 0.955 | 0.964 | 0.964 |
| 2. Chave (2001) | | 1.000 | 0.591 | 0.991 | 0.909 | 0.982 | 0.955 | 0.964 | 0.964 |
| 3. Rodríguez (1989) | | | 1.000 | 0.536 | 0.718 | 0.527 | 0.555 | 0.564 | 0.564 |
| 4. Overman (1990) | | | | 1.000 | 0.873 | 0.991 | 0.964 | 0.973 | 0.973 |
| 5. Overman *et al.* (1990) | | | | | 1.000 | 0.882 | 0.909 | 0.900 | 0.900 |
| 6. Chave (2005) | | | | | | 1.000 | 0.973 | 0.964 | 0.964 |
| 7. Anzola y Rodríguez (2001) | | | | | | | 1.000 | 0.991 | 0.991 |
| 8. Overman *et al.* (1990) | | | | | | | | 1.000 | 1.000 |
| 9. Chave *et al.* (2005) | | | | | | | | | 1.000 |
| Correlación media | 0.919 | 0.919 | 0.581 | 0.911 | 0.875 | 0.908 | 0.911 | 0.915 | 0.915 |

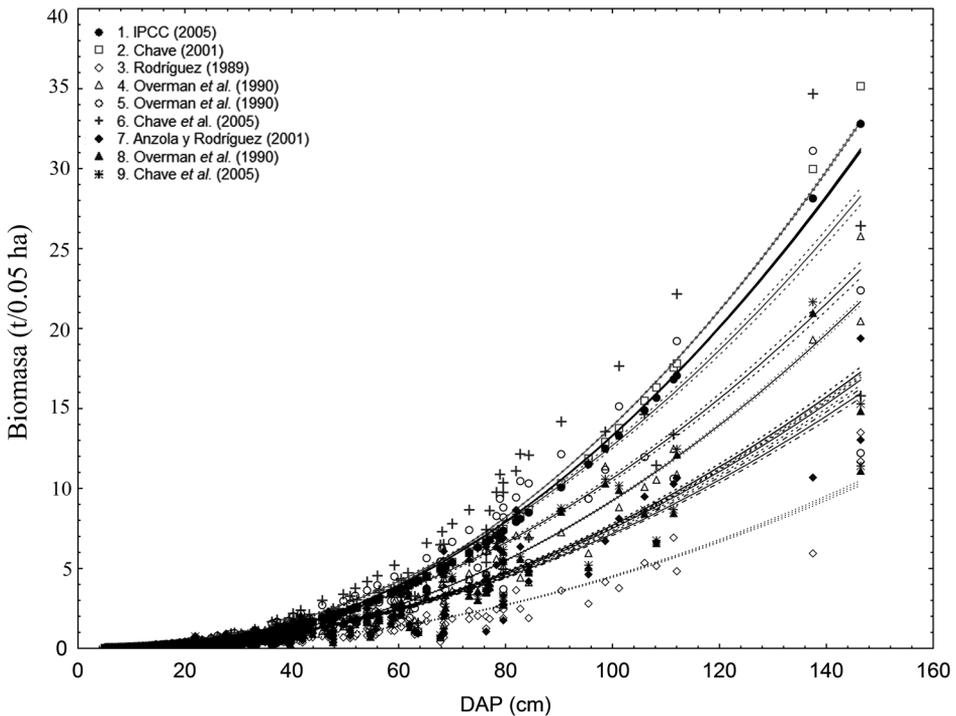

**Figura 376.** Biomasa estimada para individuos de un diámetro dado usando nueve modelos disponibles en la literatura.

florae con valores de biomasa entre 18.54 y 20.1 t/0.05ha y de carbono entre 8.59 y 9.04 t/0.05ha. En la categoría *muy bajas* (7.39-14.74) se encuentran los bosques de asociaciones Marilo laxiflorae - Pentaclethretum macrolobae, Mayno grandifoliae - Astrocaryetum malybo, Cordietum proctato - panamensis y la comunidad de *Acalypha* sp. y *Guazuma ulmifolia* cuya biomasa varía entre 7.39 y 14.09 t/0.05ha y el carbono varía entre 3.34 y 6.58 t/0.05ha.

A partir de la variación observada en la biomasa y en el carbono entre los tipos de bosque, se probó la significancia estadística de estas diferencias mediante análisis de varianza (prueba de Kruskal-Wallis). Se encontró alta variación en la densidad, área basal, altura, peso específico de la madera y biomasa entre los tipos de bosque como se evidenció con los resultados de las comparaciones de categorías y de medianas mediante las pruebas estadísticas (Tabla 168 y Figura 377). Las diferencias indican que las características de los tipos de bosque (estructura, composición florística, grado de disturbio) están relacionadas con la cantidad de biomasa almacenada en la vegetación. En la Figura 377 se observa la variación de las los parámetros de la estructura y de la biomasa entre los tipos de bosque.

En cuanto al número de individuos, se observa que algunos bosques se alejan considerablemente del promedio general dado que tienen un número alto de individuos, esta situación la muestran los bosques de las asociación Mayno grandifoliae - Astrocaryetum malybo con 157 individuos/ 0.05ha en promedio. Las asociaciones Cordietum proctato - panamensis; Tovomito weddellianae - Quercetum humboldtii y Jacarando copaiae - Pouterietum multiflorae también muestran alto número de individuos con valores de 136, 125 y 122 individuos/0.05ha. En contraste, el tipo de bosque Trichilio hirtae - Schizolobietum parahibi tiene el número de individuos más bajo (61 individuos/0.05ha).

La altura de los individuos leñosos varía entre 0.5 m y 40 m en todos los tipos de bosque, y presenta valores medios entre 6.48 m y 9.54 m. Se nota claramente que los bosques pertenecientes a las asociaciones Mayno grandifoliae - Astrocaryetum malybo, Cariniano pyriformis - Pentaplarietum doroteae, Cordietum proctato - panamensis y la

**Tabla 168.** Valores medios de las variables de la estructura, de la biomasa y del carbono en los diferentes tipos de bosque comparados mediante análisis de varianza basado en categorías (Kruskal-Wallis).

| Tipo de bosque | Densidad (Ind/0.05ha) | Área basal (m²/0.05ha) | Altura (m) | ρ (g/cm³) | Biomasa (t/0.05ha) | RCB (t/0.05ha) | Clase de RCB |
|---|---|---|---|---|---|---|---|
| Prestoeo decurrentis - Trichillietum poeppigi | 97.75 | 5.66 | 9.54 | 0.58 | 41.04±16.77 | 18.72±7.69 | *Muy alta* |
| Macrolobio ischnocalycis - Peltogynetum purpurea | 115.00 | 4.96 | 9.51 | 0.61 | 41.15±21.81 | 18.9±9.91 | *Muy alta* |
| Tovomito weddellianae - Quercetum humboldtii | 125.00 | 3.59 | 8.15 | 0.62 | 30.01±5.15 | 13.78±2.38 | *Alta* |
| Protio aracouchini - Viroletum elongatae | 109.00 | 4.17 | 8.88 | 0.57 | 34.33±11.73 | 15.66±5.53 | *Alta* |
| Trichilio hirtae- Schizolobietum parahibi | 60.67 | 3.09 | 9.44 | 0.59 | 24.06±23.26 | 10.83±10.44 | *Media* |
| Trichilio hirtae- Schizolobietum parahibi-1 | 93 | 4.95 | 8.1 | 0.64 | 46.05±34.72 | 20.65±15.66 | |
| Trichilio hirtae- Schizolobietum parahibi-2 | 44.50 | 2.17 | 10.06 | 0.57 | 13.06±4.05 | 5.92±1.84 | |
| Cappari odoratissimatis - Cavanillesietum platanifoliae | 100.33 | 3.34 | 8.29 | 0.62 | 22.45±11.69 | 10.12±5.27 | *Media* |
| Cariniano pyriformis - Pentaplarietum doroteae | 122.00 | 3.23 | 6.48 | 0.64 | 20.1±2.05 | 9.04±2 | *Baja* |
| Jacarando copaiae - Pouterietum multiflorae | 122.67 | 2.48 | 8.01 | 0.60 | 18.54±7.94 | 8.59±3.72 | *Baja* |
| Jacarando copaiae - Pouterietum multiflorae-1 | 142.00 | 2.68 | 6.82 | 0.59 | 12.52±0.76 | 5.72±0.35 | |
| Jacarando copaiae - Pouterietum multiflorae-2 | 113.00 | 2.39 | 8.61 | 0.60 | 21.545±8.48 | 10.03±3.92 | |
| Marilo laxiflorae - Pentaclethretum macrolobae | 97.00 | 2.19 | 9.28 | 0.59 | 14.09±2.25 | 6.58±1.08 | *Muy baja* |
| Mayno grandifoliae - Astrocaryetum malybo | 157.00 | 3.60 | 6.49 | 0.62 | 13.73±2.01 | 6.22±2 | *Muy baja* |
| Cordietum proctato - Cordietum panamensis | 136.00 | 2.02 | 7.13 | 0.59 | 10.51±5.88 | 4.7±2.63 | *Muy baja* |
| *Acalypha* sp. y *Guazuma ulmifolia* | 79.67 | 2.26 | 7.19 | 0.54 | 7.39±1.79 | 3.34±0.84 | *Muy baja* |
| *Estadístico H (11, N=47)* | 29.57 | 27.88 | 31.79 | 31.35 | 35.76 | 35.52 | |
| *Nivel de significancia (p)* | 0.0054 | 0.0094 | 0.0026 | 0.003 | 0.0006 | 0.0007 | |

Convenciones: Ind= número de individuos; ρ= peso específico de la madera; RCB= reserva de carbono en la biomasa aérea; Muy baja (7.39-14.74); Baja (14.74-22.09); Media (22.09-29.44); Alta (29.44-36.79) y Muy alta (36-79-44.14). Los valores corresponden a levantamientos de 0.05 ha.

comunidad de *Acalypha* sp. y *Guazuma ulmifolia* se alejan del conjunto de bosques, dado que presentan los menores valores de altura entre 6.48 y 7.19m; mientras que los bosques de las asociaciones Prestoeo decurrentis - Trichillietum poeppigi, Macrolobio ischnocalycis - Peltogynetum purpurea y Trichilio hirtae - Schizolobietum parahibi, exhiben valores ligeramente mayores al conjunto en general que varían entre 9.51, 9.54 y 9.44 m.

El peso específico de la madera tiene menor variación, pero se pueden determinar también dos grupos de bosques con niveles altos y bajos de esta variable. Es muy particular el caso del bosque de la asociación Cariniano pyriformis - Pentaplarietum doroteae, que muestra los valores más altos (0.64 g/cm$^3$), lo cual está relacionado con el alto número de individuos de la especie *Astrocaryum malybo* que presenta valores altos de peso específico de la madera (0.727 g/cm$^3$). La comunidad de *Acalypha* sp. y *Guazuma ulmifolia* y la asociación Protio aracouchini - Viroletum elongatae mostraron los valores más bajos de esta variable (0.54 y 0.57 g/cm$^3$, respectivamente).

Con relación al área basal, se diferencian claramente los bosques de las asociaciones Prestoeo decurrentis - Trichillietum poeppigi, Macrolobio ischnocalycis - Peltogynetum purpurea, Protio aracouchini - Viroletum elongatae y Tovomito weddellianae - Quercetum humboldtii, los cuales presentan los valores mayores de área basal entre 3.59 y 5.66 m$^2$/0.05ha. En contraste, los bosques de los tipos Marilo laxiflorae - Pentaclethretum macrolobae, Cordietum proctato - Cordietum panamensis y *Acalypha* sp. y *Guazuma ulmifolia* presentaron la menor área basal con valores comprendidos entre 2.02 y 2.19 m$^2$/0.05ha.

La biomasa muestra alta variación como resultado de la variabilidad conjunta de los parámetros de la estructura antes descritos. Los bosques con *altas* reservas de biomasa y carbono tienen valores altos de área basal y altura, este es el caso para las asociaciones Prestoeo decurrentis - Trichillietum poeppigi, Macrolobio ischnocalycis - Peltogynetum purpurea y Protio aracouchini - Viroletum elongatae; la asociación Tovomito weddellianae - Quercetum humboldtii tiene valores relativamente altos de área basal y número de individuos. Los bosques con reservas de biomasa y carbono *muy bajas* tienen valores bajos de área basal y altura como en el caso de las asociaciones Cordietum proctato - Cordietum panamensis y Marilo laxiflorae - Pentaclethretum macrolobae; presentan alto número de individuos como en el caso de Mayno grandifoliae - Astrocaryetum malybo y Cordietum proctato - Cordietum panamensis y bajos valores de área basal, altura, peso específico de la madera y número de individuos en el caso de la comunidad de *Acalypha* sp. y *Guazuma ulmifolia.*

Se probó la significancia estadística de las diferencias en las variables de la estructura y de la biomasa entre los tipos de bosque, mediante comparación de categorías usando la prueba Kruskal-Wallis por clase diamétrica, para determinar cuáles clases varían. Se encontraron diferencias significativas en la densidad, área basal, altura y biomasa en las clases diamétricas (2.5-70 cm). La ausencia de árboles de gran porte en algunos tipos de bosque no permitió determinar las diferencias entre las variables en la clase diamétrica mayor [>70 cm]. (Tabla 169).

En la tabla 170 se muestra la distribución de las variables que se asocian con la estructura de la vegetación (altura, área basal, número de individuos), el peso específico de la madera, la biomasa y el carbono por clase diamétrica en los diferentes tipos de bosque. Los datos de altura corresponden a valores mínimo y máximo; el peso específico es el valor promedio por asociación. Los valores de área basal, densidad, biomasa y carbono corresponden al promedio de cada asociación el cual se obtuvo al sumar todos los individuos de cada levantamiento y promediar los levantamientos de cada asociación. Además se enuncia la participación en porcentaje de las variables en cada clase de diámetro, para cada tipo de bosque. Se resaltan las particularidades.

Esta distribución indica variación de la estructura considerable a lo largo del gradiente Sur – Noroccidente de Córdoba, que conduce a su vez a una alta variación en la biomasa aérea de los bosques asociados a dicho gradiente. La estructura de la vegetación entre los tipos de bosque varió en términos de distribución de clases diamétricas, las reservas de biomasa y el carbono. La diferencia entre presencia y ausencia de individuos de gran porte (clases diamétricas y estratos superiores) entre los diferentes tipos de bosque está relacionada

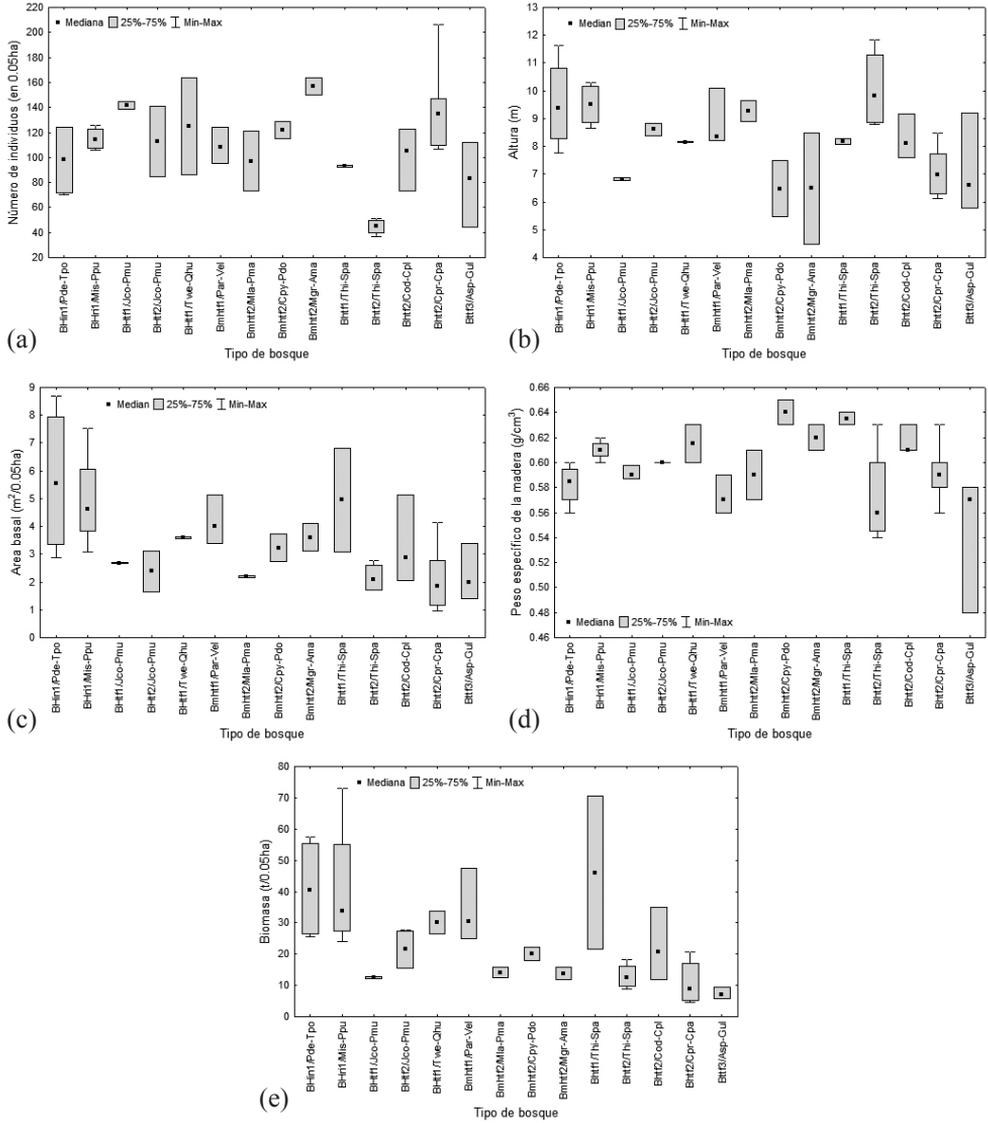

**Figura 377.** Variación de parámetros de la estructura y de biomasa en los tipos de bosque.
(a) Número de individuos, (b) Altura, (c) Área basal, (d) Peso específico de la madera y (e) Biomasa.

**Tabla 169.** Comparaciones de las variables de la estructura y la biomasa por clase diamétrica entre tipos de bosque.

Análisis de varianza basado en categorías (Kluskal-Wallis). H: estadístico de la prueba, *p*: nivel de significancia estadística; ρ: peso específico de la madera. ns: no significativa.

| Clase diamétrica (cm) | Variable | Prueba de Kruskal-Wallis | |
|---|---|---|---|
| I (2.5-10) | Altura (m) | H ( 11, N= 3120) =283.9125 p =0.000 | |
| | ρ (g/cm$^3$) | H ( 11, N= 3120) =235.6824 p =0.000 | |
| | Área basal (m$^2$) | H ( 11, N= 3120) =334.6173 p =0.000 | |
| | Número de individuos | H ( 11, N= 47) =27.32425 p =.0041 | |
| | Biomasa (kg) | H ( 11, N= 3120) =204.3114 p =0.000 | |
| II (10-30) | Altura (m) | H ( 11, N= 1206) =81.61550 p =.0000 | |
| | ρ (g/cm$^3$) | H ( 11, N= 1206) =92.78781 p =.0000 | |
| | Área basal (m$^2$) | H ( 11, N= 1206) =16.19571 p =.1340 | ns |
| | Número de individuos | H ( 11, N= 47) =24.35153 p =.0113 | |
| | Biomasa (kg) | H ( 11, N= 1206) =50.99291 p =.0000 | |
| III (30-50) | Altura (m) | H ( 11, N= 222) =45.57259 p =.0000 | |
| | ρ (g/cm$^3$) | H ( 11, N= 222) =27.96300 p =.0033 | |
| | Área basal (m$^2$) | H ( 11, N= 222) =23.07745 p =.0172 | |
| | Número de individuos | H ( 11, N= 47) =8.796266 p =.6407 | ns |
| | Biomasa (kg) | H ( 11, N= 222) =47.56937 p =.0000 | |
| IV (50-70) | Altura (m) | H ( 11, N= 61) =20.17633 p =.0430 | |
| | ρ (g/cm$^3$) | H ( 11, N= 61) =31.62133 p =.0009 | |
| | Área basal (m$^2$) | H ( 11, N= 61) =6.122159 p =.8651 | ns |
| | Número de individuos | H ( 11, N= 47) =19.69893 p =.0496 | ns |
| | Biomasa (kg) | H ( 11, N= 61) =29.61410 p =.0018 | ns |
| V (70-100) | Altura (m) | H ( 11, N= 37) =17.84457 p =.0853 | ns |
| | ρ (g/cm$^3$) | H ( 11, N= 37) =18.21889 p =.0766 | ns |
| | Área basal (m$^2$) | H ( 11, N= 37) =13.87062 p =.2402 | ns |
| | Número de individuos | H ( 11, N= 47) =21.87487 p =.0254 | |
| | Biomasa (kg) | H ( 11, N= 37) =18.72262 p =.0663 | ns |

con la variación en las reservas de biomasa y carbono observadas. Por ejemplo la comunidad de *Acalypha* sp. y *Guazuma ulmifolia* presenta el menor valor de biomasa (7.39 t/0.05 ha), lo cual está en relación con la ausencia de árboles mayores a 100 cm de diámetro y la distribución de la biomasa principalmente en las clases diamétricas inferiores; la clase II (10-30) contiene la mayor proporción de biomasa con un valor de 38.97%. En contraste la asociación Trichilio hirtae- Schizolobietum parahibi_1 se caracteriza por presentar altos valores de biomasa (41.15 t/0.05 ha), la cual se concentra principalmente en la clase diamétrica VI (>100) con un promedio de 25.83 t/0.05 ha (62.78%).

En los tipos de bosque se observaron diferencias en la distribución de clases de diámetro, las asociaciones Jacarando copaiae - Pouterietum multiflorae_1, Jacarando copaiae - Pouterietum multiflorae_2, Marilo laxiflorae - Pentaclethretum macrolobae, Mayno grandifoliae - Astrocaryetum malybo, Cariniano pyriformis - Pentaplarietum doroteae y *Acalypha* sp. y *Guazuma ulmifolia* no presentan individuos en la clase VI (>100cm). A diferencia de éstas, las asociaciones Prestoeo decurrentis - Trichillietum poeppigi y Macrolobio ischnocalycis - Peltogynetum purpurea presentan el número más alto de individuos promedio en esta clase diamétrica (2 individuos en 0.05ha); estos resultados muestran que estas asociaciones presentan la mayor área basal y biomasa en esta clase, en comparación con las demás asociaciones.

Las asociaciones Tovomito weddellianae - Quercetum humboldtii, Protio aracouchini - Viroletum elongatae y Cariniano pyriformis - Pentaplarietum doroteae presentan los mayores valores de área basal en las clases IV (50-70cm) y V (70-100cm), que conduce a que presenten el segundo valor de biomasa más alto. En contraste las asociaciones Mayno grandifoliae - Astrocaryetum malybo, Marilo laxiflorae - Pentaclethretum macrolobae y Jacarando copaiae - Pouterietum multiflorae tienen los valores más altos de área basal en la clase diamétrica II (10-30cm).

**Tabla 170.** Distribución de las variables de la estructura de la vegetación, biomasa y carbono por clase diamétrica en los tipos de bosques caracterizados en el sector Sur y Noroccidente del departamento de Córdoba.

| Asociación | Variable | Clase diamétrica | | | | | | |
|---|---|---|---|---|---|---|---|---|
| | | I (2.5-10) | II (10-30) | III (30-50) | IV (50-70) | V (70-100) | VI (>100) | Total |
| Prestoeo decurrentis - Trichillietum poeppigi | Altura (m) | 2-20 | 4-30 | 4-30 | 10-25 | 20-35 | 27-35 | 2-35 |
| | Peso específico | 0.59 | 0.57 | 0.57 | 0.43 | 0.56 | 0.49 | 0.58 |
| | Densidad | 65 | 23.5 | 4.25 | 2 | 1.5 | 1.5 | 97.75 |
| | Densidad relativa (%) | 66.5 | 24.04 | 4.35 | 2.05 | 1.53 | 1.53 | 100 |
| | Área basal (m²/0.05ha) | 0.17 | 0.55 | 0.5 | 0.58 | 0.87 | 2.99 | 5.65 |
| | Área basal relativa (%) | 3.03 | 9.65 | 8.83 | 10.28 | 15.3 | 52.9 | 100 |
| | Biomasa (t/0.05ha) | 0.44 | 2.8 | 3.46 | 3.16 | 9 | 22.18 | 41.03 |
| | Biomasa relativa (%) | 1.06 | 6.83 | 8.42 | 7.69 | 21.93 | 54.06 | 100 |
| | Carbono (t/0.05ha) | 0.18 | 1.29 | 1.61 | 1.45 | 4.07 | 10.12 | 18.72 |
| | Carbono relativo (%) | 0.96 | 6.92 | 8.59 | 7.75 | 21.74 | 54.04 | 100 |
| Macrolobio ischnocalycis - Peltogynetum purpurea | Altura (m) | 2-22 | 6-20 | 5-25 | 20-22 | 20-28 | 26-31 | 2-31 |
| | Peso específico | 0.61 | 0.61 | 0.6 | 0.64 | 0.8 | 0.66 | 0.61 |
| | Densidad | 78.25 | 28.5 | 4.75 | 1 | 0.75 | 1.75 | 115 |
| | Densidad relativa (%) | 68.04 | 24.78 | 4.13 | 0.87 | 0.65 | 1.52 | 100 |
| | Área basal (m²/0.05ha) | 0.21 | 0.64 | 0.58 | 0.24 | 0.38 | 2.91 | 4.96 |
| | Área basal relativa (%) | 4.2 | 12.83 | 11.75 | 4.81 | 7.61 | 58.79 | 100 |
| | Biomasa (t/0.05ha) | 0.64 | 3.52 | 4.34 | 2 | 4.82 | 25.83 | 41.15 |
| | Biomasa relativa (%) | 1.55 | 8.55 | 10.55 | 4.85 | 11.72 | 62.78 | 100 |
| | Carbono (t/0.05ha) | 0.27 | 1.63 | 2.03 | 0.92 | 2.21 | 11.84 | 18.9 |
| | Carbono relativo (%) | 1.4 | 8.63 | 10.74 | 4.86 | 11.71 | 62.65 | 100 |
| Tovomito weddellianae - Quercetum humboldtii | Altura (m) | 1.8-12 | 5-20 | 10-27 | 18-27 | 18-27 | 0 | 1.8-27 |
| | Peso específico | 0.6 | 0.63 | 0.6 | 0.78 | 0.72 | 0 | 0.61 |
| | Densidad | 77 | 36 | 7 | 3 | 2 | 0 | 125 |
| | Densidad relativa (%) | 61.6 | 28.8 | 5.6 | 2.4 | 1.6 | 0 | 100 |
| | Área basal (m²/0.05ha) | 0.21 | 0.81 | 0.83 | 0.83 | 0.9 | 0 | 3.59 |
| | Área basal relativa (%) | 5.96 | 22.68 | 23.1 | 23.14 | 25.12 | 0 | 100 |
| | Biomasa (t/0.05ha) | 0.48 | 4.19 | 6.43 | 9.71 | 9.19 | 0 | 30 |
| | Biomasa relativa (%) | 1.59 | 13.98 | 21.44 | 32.36 | 30.64 | 0 | 100 |
| | Carbono (t/0.05ha) | 0.19 | 1.94 | 2.97 | 4.46 | 4.23 | 0 | 13.79 |
| | Carbono relativo (%) | 1.37 | 14.07 | 21.52 | 32.39 | 30.65 | 0 | 100 |
| Protio aracouchini - Viroletum elongatae | Altura (m) | 2-12 | 3-25 | 4-25 | 20-30 | 24-30 | 30-35 | 2-35 |
| | Peso específico | 0.58 | 0.56 | 0.54 | 0.59 | 0.58 | 0.58 | 0.57 |
| | Densidad | 66.33 | 30 | 7.33 | 3 | 1.67 | 0.67 | 109 |
| | Densidad relativa (%) | 60.86 | 27.52 | 6.73 | 2.75 | 1.53 | 0.61 | 100 |
| | Área basal (m²/0.05ha) | 0.16 | 0.76 | 0.75 | 0.89 | 1 | 0.61 | 4.17 |
| | Área basal relativa (%) | 3.9 | 18.14 | 18.09 | 21.32 | 23.9 | 14.65 | 100 |
| | Biomasa (t/0.05ha) | 0.37 | 3.53 | 5.01 | 8.18 | 9.85 | 7.4 | 34.33 |
| | Biomasa relativa (%) | 1.08 | 10.28 | 14.6 | 23.82 | 28.69 | 21.54 | 100 |
| | Carbono (t/0.05ha) | 0.14 | 1.63 | 2.29 | 3.78 | 4.44 | 3.38 | 15.66 |
| | Carbono relativo (%) | 0.91 | 10.39 | 14.62 | 24.13 | 28.36 | 21.59 | 100 |
| Jacarando copaiae - Pouterietum multiflorae_1 | Altura (m) | 1.7-10 | 4-25 | 3.5-25 | 5-16 | 0 | 0 | 1.7-25 |
| | Peso específico | 0.59 | 0.6 | 0.5 | 0.63 | 0.6 | 0 | 0.59 |
| | Densidad | 95 | 39 | 6 | 2 | 0 | 0 | 142 |
| | Densidad relativa (%) | 66.9 | 27.46 | 4.23 | 1.41 | 0 | 0 | 100 |
| | Área basal (m²/0.05ha) | 0.2 | 1.06 | 0.79 | 0.63 | 0 | 0 | 2.68 |
| | Área basal relativa (%) | 7.52 | 39.44 | 29.56 | 23.49 | 0 | 0 | 100 |
| | Biomasa (t/0.05ha) | 0.38 | 4.96 | 4.79 | 2.38 | 0 | 0 | 12.52 |
| | Biomasa relativa (%) | 3.03 | 39.63 | 38.3 | 19.04 | 0 | 0 | 100 |
| | Carbono (t/0.05ha) | 0.13 | 2.28 | 2.21 | 1.1 | 0 | 0 | 5.72 |
| | Carbono relativo (%) | 2.34 | 39.81 | 38.67 | 19.18 | 0 | 0 | 100 |

**Continuación Tabla 170.** Distribución de las variables de la estructura de la vegetación, biomasa y carbono por clase diamétrica en los tipos de bosques caracterizados en el sector Sur y Noroccidente del departamento de Córdoba.

| Asociación | Variable | Clase diamétrica | | | | | | |
|---|---|---|---|---|---|---|---|---|
| | | I (2.5-10) | II (10-30) | III (30-50) | IV (50-70) | V (70-100) | VI (>100) | Total |
| Jacarando copaiae - Pouterietum multiflorae_2 | Altura (m) | 0.5-15 | 2-39 | 14-30 | 24 | 30-32 | 0 | 0.5-39 |
| | Peso específico | 0.61 | 0.6 | 0.57 | 0.47 | 0.68 | 0 | 0.6 |
| | Densidad | 76 | 29.5 | 5.5 | 0.5 | 1.5 | 0 | 113 |
| | Densidad relativa (%) | 67.26 | 26.11 | 4.87 | 0.44 | 1.33 | 0 | 100 |
| | Área basal (m$^2$/0.05ha) | 0.17 | 0.74 | 0.68 | 0.11 | 0.68 | 0 | 2.38 |
| | Área basal relativa (%) | 7.03 | 31.14 | 28.55 | 4.83 | 28.46 | 0 | 100 |
| | Biomasa (t/0.05ha) | 0.38 | 4.9 | 6.03 | 0.83 | 9.4 | 0 | 21.54 |
| | Biomasa relativa (%) | 1.77 | 22.76 | 27.99 | 3.86 | 43.63 | 0 | 100 |
| | Carbono (t/0.05ha) | 0.14 | 2.3 | 2.81 | 0.38 | 4.39 | 0 | 10.03 |
| | Carbono relativo (%) | 1.44 | 22.9 | 28.05 | 3.82 | 43.79 | 0 | 100 |
| Marilo laxiflorae - Pentaclethretum macrolobae | Altura (m) | 2.5-17 | 4-25 | 15-25 | 27 | 20 | 0 | 2.5-27 |
| | Peso específico | 0.59 | 0.57 | 0.58 | 0.64 | 0.65 | 0 | 0.59 |
| | Densidad | 54 | 38 | 3.5 | 0.5 | 1 | 0 | 97 |
| | Densidad relativa (%) | 55.67 | 39.18 | 3.61 | 0.52 | 1.03 | 0 | 100 |
| | Área basal (m$^2$/0.05ha) | 0.17 | 1.06 | 0.35 | 0.15 | 0.46 | 0 | 2.19 |
| | Área basal relativa (%) | 7.71 | 48.58 | 15.77 | 6.84 | 21.11 | 0 | 100 |
| | Biomasa (t/0.05ha) | 0.42 | 5.57 | 2.56 | 1.67 | 3.87 | 0 | 14.09 |
| | Biomasa relativa (%) | 2.97 | 39.52 | 18.14 | 11.88 | 27.49 | 0 | 100 |
| | Carbono (t/0.05ha) | 0.18 | 2.58 | 1.19 | 0.79 | 1.84 | 0 | 6.57 |
| | Carbono relativo (%) | 2.7 | 39.26 | 18.12 | 11.96 | 27.96 | 0 | 100 |
| Mayno grandifoliae - Astrocaryetum malybo | Altura (m) | 1.7-8 | 3-18 | 12-20 | 17-20 | 25-25 | 0 | 1.7-25 |
| | Peso específico | 0.67 | 0.58 | 0.52 | 0.42 | 0.34 | 0 | 0.62 |
| | Densidad | 84 | 65 | 5 | 2 | 1 | 0 | 157 |
| | Densidad relativa (%) | 53.5 | 41.4 | 3.18 | 1.27 | 0.64 | 0 | 100 |
| | Área basal (m$^2$/0.05ha) | 0.36 | 1.75 | 0.45 | 0.55 | 0.5 | 0 | 3.6 |
| | Área basal relativa (%) | 9.87 | 48.6 | 12.53 | 15.18 | 13.82 | 0 | 100 |
| | Biomasa (t/0.05ha) | 0.62 | 5.07 | 2.57 | 2.7 | 2.77 | 0 | 13.73 |
| | Biomasa relativa (%) | 4.49 | 36.97 | 18.69 | 19.64 | 20.21 | 0 | 100 |
| | Carbono (t/0.05ha) | 0.27 | 2.32 | 1.19 | 1.21 | 1.22 | 0 | 6.22 |
| | Carbono relativo (%) | 4.39 | 37.34 | 19.08 | 19.54 | 19.65 | 0 | 100 |
| Cariniano pyriformis - Pentaplarietum doroteae | Altura (m) | 1.3-10 | 3-20 | 18-20 | 18-28 | 25-30 | 0 | 1.3-30 |
| | Peso específico | 0.67 | 0.61 | 0.54 | 0.57 | 0.41 | 0 | 0.64 |
| | Densidad | 75 | 39 | 3 | 3 | 2 | 0 | 122 |
| | Densidad relativa (%) | 61.48 | 31.97 | 2.46 | 2.46 | 1.64 | 0 | 100 |
| | Área basal (m$^2$/0.05ha) | 0.25 | 0.78 | 0.28 | 0.86 | 1.06 | 0 | 3.23 |
| | Área basal relativa (%) | 7.72 | 24.22 | 8.69 | 26.63 | 32.74 | 0 | 100 |
| | Biomasa (t/0.05ha) | 0.37 | 3.04 | 1.81 | 7.09 | 7.8 | 0 | 20.1 |
| | Biomasa relativa (%) | 1.84 | 15.11 | 8.99 | 35.25 | 38.81 | 0 | 100 |
| | Carbono (t/0.05ha) | 0.16 | 1.37 | 0.83 | 3.15 | 3.53 | 0 | 9.04 |
| | Carbono relativo (%) | 1.78 | 15.1 | 9.21 | 34.86 | 39.05 | 0 | 100 |
| Trichilio hirtae- Schizolobietum parahibi_1 | Altura (m) | 1-12 | 5-20 | 10-25 | 10-27 | 18-28 | 30-40 | 1-40 |
| | Peso específico | 0.71 | 0.63 | 0.54 | 0.51 | 0.9 | 0.41 | 0.66 |
| | Densidad | 66 | 13 | 9.5 | 2.5 | 1 | 1 | 93 |
| | Densidad relativa (%) | 70.97 | 13.98 | 10.22 | 2.69 | 1.08 | 1.08 | 100 |
| | Área basal (m$^2$/0.05ha) | 0.18 | 0.37 | 0.93 | 0.7 | 0.51 | 2.26 | 4.95 |
| | Área basal relativa (%) | 3.67 | 7.48 | 18.8 | 14.06 | 10.38 | 45.6 | 100 |
| | Biomasa (t/0.05ha) | 0.47 | 2.09 | 5.81 | 4.89 | 4.84 | 27.95 | 46.05 |
| | Biomasa relativa (%) | 1.02 | 4.53 | 12.62 | 10.61 | 10.52 | 60.7 | 100 |
| | Carbono (t/0.05ha) | 0.19 | 0.93 | 2.61 | 2.17 | 2.16 | 12.57 | 20.64 |
| | Carbono relativo (%) | 0.92 | 4.51 | 12.65 | 10.53 | 10.47 | 60.92 | 100 |

**Continuación Tabla 170.** Distribución de las variables de la estructura de la vegetación, biomasa y carbono por clase diamétrica en los tipos de bosques caracterizados en el sector Sur y Noroccidente del departamento de Córdoba.

| Asociación | Variable | Clase diamétrica | | | | | | |
|---|---|---|---|---|---|---|---|---|
| | | I (2.5-10) | II (10-30) | III (30-50) | IV (50-70) | V (70-100) | VI (>100) | Total |
| Trichilio hirtae-Schizolobietum parahibi_2 | Altura (m) | 1.5-13 | 4-22 | 12-24 | 15-27 | 17-26 | 28 | 1.5-28 |
| | Peso específico | 0.61 | 0.56 | 0.52 | 0.56 | 0.4 | 0.6 | 0.57 |
| | Densidad | 15.25 | 21.25 | 5.5 | 1.5 | 0.75 | 0.25 | 44.5 |
| | Densidad relativa (%) | 34.27 | 47.75 | 12.36 | 3.37 | 1.69 | 0.56 | 100 |
| | Área basal (m$^2$/0.05ha) | 0.04 | 0.55 | 0.59 | 0.41 | 0.37 | 0.22 | 2.17 |
| | Área basal relativa (%) | 1.75 | 25.28 | 27.03 | 19.09 | 16.87 | 9.99 | 100 |
| | Biomasa (t/0.05ha) | 0.09 | 2.15 | 3.25 | 3.11 | 2.11 | 2.36 | 13.06 |
| | Biomasa relativa (%) | 0.66 | 16.43 | 24.89 | 23.82 | 16.16 | 18.05 | 100 |
| | Carbono (t/0.05ha) | 0.03 | 0.98 | 1.48 | 1.41 | 0.95 | 1.07 | 5.91 |
| | Carbono relativo (%) | 0.59 | 16.5 | 24.95 | 23.85 | 16.06 | 18.05 | 100 |
| Cordietum proctato - Cordietum panamensis | Altura (m) | 0.5-28 | 1.6-24 | 1.6-30 | 15-26 | 20-30 | 26-28 | 0.5-30 |
| | Peso específico | 0.6 | 0.56 | 0.56 | 0.43 | 0.51 | 0.5 | 0.59 |
| | Densidad | 105.18 | 25.45 | 4.27 | 0.73 | 0.18 | 0.18 | 136 |
| | Densidad relativa (%) | 77.34 | 18.72 | 3.14 | 0.53 | 0.13 | 0.13 | 100 |
| | Área basal (m$^2$/0.05ha) | 0.19 | 0.64 | 0.51 | 0.21 | 0.1 | 0.37 | 2.02 |
| | Área basal relativa (%) | 9.38 | 31.79 | 25.41 | 10.5 | 4.74 | 18.18 | 100 |
| | Biomasa (t/0.05ha) | 0.58 | 3.16 | 2.93 | 1.17 | 0.7 | 1.98 | 10.51 |
| | Biomasa relativa (%) | 5.47 | 30.03 | 27.88 | 11.12 | 6.66 | 18.83 | 100 |
| | Carbono (t/0.05ha) | 0.22 | 1.43 | 1.32 | 0.53 | 0.32 | 0.88 | 4.7 |
| | Carbono relativo (%) | 4.59 | 30.5 | 28.16 | 11.25 | 6.72 | 18.78 | 100 |
| Cappari odoratissimatis - Cavanillesietum platanifoliae | Altura (m) | 1-18 | 2.5-25 | 14-28 | 15 | 25-28 | 20-28 | 1-28 |
| | Peso específico | 0.62 | 0.63 | 0.63 | 0.58 | 0.45 | 0.53 | 0.62 |
| | Densidad | 67.33 | 25.67 | 5.33 | 0.33 | 0.67 | 1 | 100.33 |
| | Densidad relativa (%) | 67.11 | 25.58 | 5.32 | 0.33 | 0.66 | 1 | 100 |
| | Área basal (m$^2$/0.05ha) | 0.15 | 0.64 | 0.66 | 0.08 | 0.49 | 1.32 | 3.34 |
| | Área basal relativa (%) | 4.44 | 19.27 | 19.83 | 2.32 | 14.53 | 39.61 | 100 |
| | Biomasa (t/0.05ha) | 0.49 | 3.9 | 5.31 | 0.41 | 3.1 | 9.25 | 22.45 |
| | Biomasa relativa (%) | 2.18 | 17.38 | 23.63 | 1.81 | 13.79 | 41.2 | 100 |
| | Carbono (t/0.05ha) | 0.19 | 1.76 | 2.4 | 0.18 | 1.4 | 4.18 | 10.12 |
| | Carbono relativo (%) | 1.92 | 17.39 | 23.77 | 1.82 | 13.79 | 41.31 | 100 |
| Acalypha sp. y Guazuma ulmifolia | Altura (m) | 1.5-16 | 2.5-23 | 6-16 | 7-18 | 8 | 0 | 1.5-23 |
| | Peso específico | 0.59 | 0.49 | 0.44 | 0.5 | 0.54 | 0 | 0.54 |
| | Densidad | 42.67 | 30 | 4.67 | 1.67 | 0.67 | 0 | 79.67 |
| | Densidad relativa (%) | 53.56 | 37.66 | 5.86 | 2.09 | 0.84 | 0 | 100 |
| | Área basal (m$^2$/0.05ha) | 0.11 | 0.74 | 0.56 | 0.43 | 0.43 | 0 | 2.26 |
| | Área basal relativa (%) | 4.85 | 32.68 | 24.6 | 19.01 | 18.86 | 0 | 100 |
| | Biomasa (t/0.05ha) | 0.28 | 2.88 | 1.75 | 1.38 | 1.09 | 0 | 7.39 |
| | Biomasa relativa (%) | 3.83 | 38.97 | 23.74 | 18.73 | 14.73 | 0 | 100 |
| | Carbono (t/0.05ha) | 0.11 | 1.3 | 0.8 | 0.64 | 0.5 | 0 | 3.34 |
| | Carbono relativo (%) | 3.37 | 38.79 | 23.81 | 19.07 | 14.96 | 0 | 100 |

La alta variabilidad estructural encontrada en este trabajo se asemeja a otros estudios, como el realizado por Sierra *et al* (2007) en la región de Porce, Antioquia, Colombia, en el cual se cuantificaron las reservas de carbono en un paisaje heterogéneo compuesto por bosques primarios y secundarios. Ellos encontraron alta variabilidad en la estructura de los bosques y mencionan que el área basal media en los bosques primarios fue tres veces mayor (36.85 m$^2$/ha) que la de bosques secundarios (12.92m$^2$/ha). Esta variación se relaciona también con la historia de uso del suelo y otros factores como las características físicas y químicas del suelo y la pendiente que al interactuar producen alta variación en la cobertura boscosa de esa región.

Lindner (2010) en bosques húmedos del Atlántico en Brasil, también encontró alta variación en la biomasa, que estaba relacionada con la presencia de árboles grandes en varios levantamientos de vegetación. En dicho estudio, los árboles con DAP>30cm representaron solo el 6% de los indi-viduos muestreados, pero contribuyeron al 72% de la biomasa estimada.

**Distribución de la biomasa y el carbono por clase diamétrica**

La distribución de la biomasa y el carbono por clase diamétrica (DAP) varió notablemente entre los tipos de bosque. Los bosques que pertenecen a la misma asociación pero que se distribuyen en localidades diferentes se presentan por separado, diferenciados por un número, el cual indica su nivel de intervención (1: poco intervenido, 2: intervención media-alta), esto sucede con las asociaciones Trichilio hirtae- Schizolobietum parahibi y Jacarando copaiae-Pouterietum multiflorae. En la Figura 378 se presenta la distribución de la biomasa y de carbono por clase diamétrica para cada tipo de bosque. Los bosques que presentaron una tendencia similar fueron agrupados y etiquetados con una letra, obteniéndose seis grupos (A-F).

El primer grupo (A) se caracteriza por los bosques que presentan la mayor acumulación de la biomasa y del carbono en los árboles de clases diamétricas mayores (clases X-XIII, DAP>100cm). Este patrón lo exhiben las asociaciones Prestoeo decurrentis-Trichillietum poeppigi, Macrolobio ischnocalycis-Peltogynetum purpurea y Trichilio hirtae-Schizolobietum parahibi_1. En la primera asociación, la biomasa y el carbono se distribuyen en su mayoría (69%) en tres clases diamétricas X(90-100), XI(100-150) y XII(150-200), lo cual corresponde a 28.32 t/0.05ha; en la segunda asociación la biomasa y el carbono se concentran principalmente en dos clases (25.83 t/0.05ha que equivalen al 62.77% en las clases XI(100-150) y XII(150-200)). Una tendencia similar muestra la tercera asociación con una acumulación mayor de biomasa y carbono en una clase (24.11 t/0.05ha que corresponden a 54.34% del total de la asociación en la clase XIII(>200).

El grupo B lo conforman los bosques de las asociaciones Tovomito weddellianae- Quercetum humboldtii y Protio aracouchini-Viroletum elongatae, los cuales presentan una distribución completa, es decir poseen individuos en todas las clases diamétricas. En la primera asociación hay una acumulación importante de biomasa y carbono en las clases VIII(70-80) con un 22.16% (6.65t/0.05ha), las clases VI(50-60), VII(60-70) y IV(30-40) tienen una participación considerable en la acumulación de biomasa en este bosque contribuyendo con un 16.46% (4.94 t/0.05ha), 15.89% (4.77 t/0.05ha) y 13.97% (4.19t/0.05ha), respectivamente; no se observa sin embargo una diferenciación tan marcada de pocas clases con respecto a las demás, como sucede en los bosques del grupo A. En la segunda asociación, las clases VII(60-70), XI(100-150) y VIII(70-80) concentran una alta proporción de biomasa y carbono con valores de 21.54, 21.27 y 11.40%, respectivamente (7.30, 7.39 y 3.91 t/0.05ha).

El grupo C lo conforman los bosques de la asociación Trichilio hirtae-Schizolobietum parahibi_2, los cuales presentan la mayor acumulación de biomasa en la clase XI(100-150) con un 18% (2.36 t/0.05ha), seguido por las clases IV, V y VI (30-60) con 12.90% (1.69 t/0.05ha), 12.28% (1.57 t/0.05ha), y 11.51% (1.51 t/0.05ha), respectivamente. La biomasa se reparte en cantidades similares en las clases III-VIII (20-80).

El grupo D se caracteriza por presentar distribución incompleta, es decir en algunas clases diamétricas no hay individuos. La biomasa se concentra en las clases menores (III-V: 20-50) y mayores (VIII-XII: 80-200). Se distinguen dos grupos separados por la ausencia de árboles en las clases medias, con cantidades de biomasa y carbono similares en ambos grupos. Esta tendencia la presentan los bosques de las asociaciones Cappari odoratissimatis - Cavanillesietum platanifoliae; Cariniano pyriformis -Pentaplarietum doroteae; Jacarando copaiae-Pouterietum multiflorae_2; Mayno grandifoliae - Astrocaryetum malybo y Marilo laxiflorae - Pentaclethretum macrolobae.

El grupo E lo constituyen los bosques de la asociación Jacarando copaiae-Pouterietum multiflorae_1 que presentan una distribución completa (sin interrupciones), pero no contiene individuos con diámetros mayores a 80 cm, contiene la mayor cantidad de biomasa y carbono en las clases V(40-50) y III(20-30) con 25.68% (3.09 t/0.05ha) y 24.04% (3.01 t/0.05ha). La cantidad de biomasa en las clases II(10-20), IV(30-40) y VI(50-60) es muy similar: 15.57% (1.95t/0.05ha), 13.65 (1.71 t/0.05ha) y 12.46% (1.56 t/0.05ha).

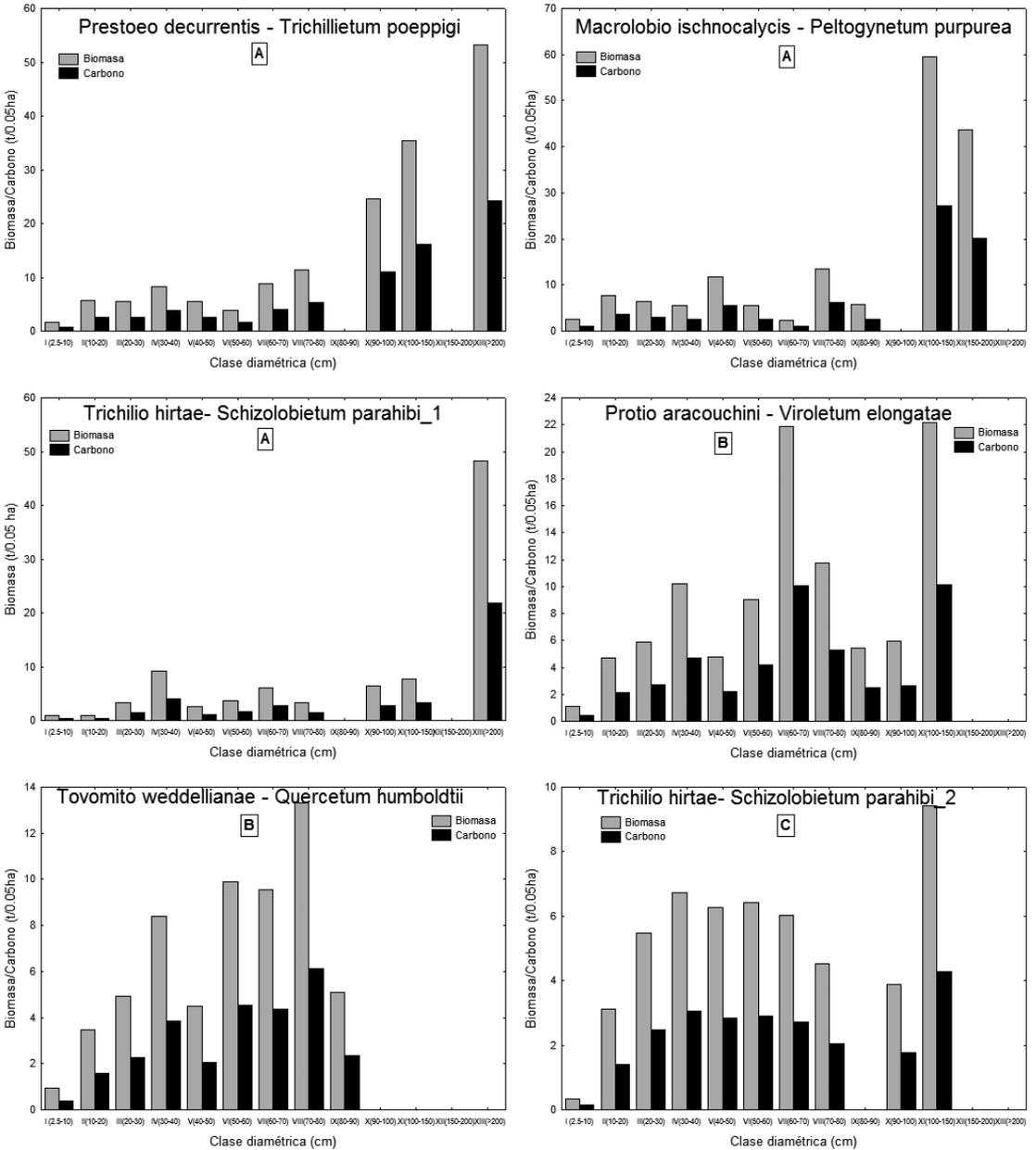

**Figura 378.** Distribución de la biomasa aérea por clase diamétrica de los individuos con DAP > a 2.5 cm de en los diferentes tipos de bosque de localidades del Sur y Noroccidente de Córdoba, Colombia. Las letras en los cuadrados indican los grupos de asociaciones.

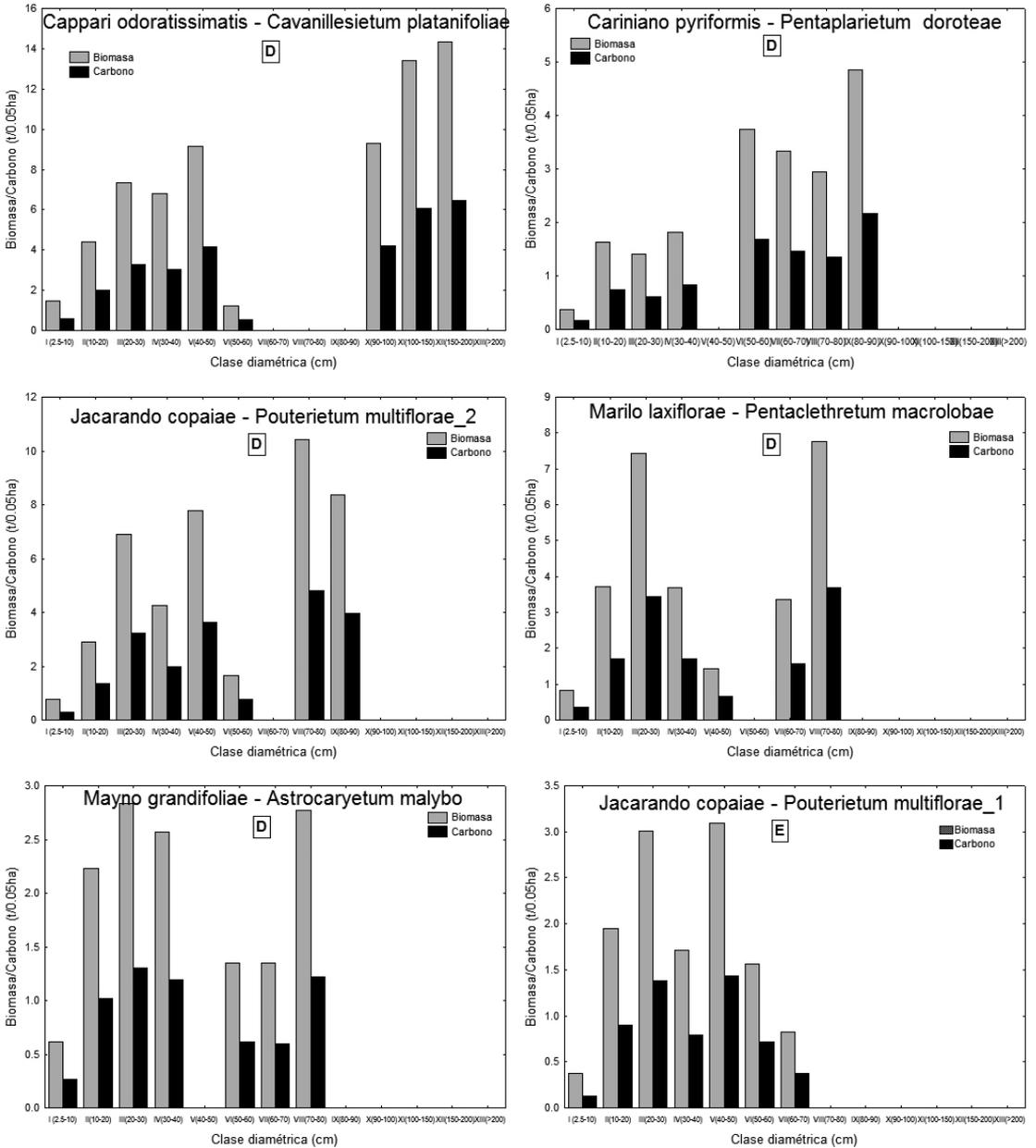

**Continuación Figura 378.** Distribución de la biomasa aérea por clase diamétrica de los individuos con DAP > a 2.5 cm de en los diferentes tipos de bosque de localidades del Sur y Noroccidente de Córdoba, Colombia. Las letras en los cuadrados indican los grupos de asociaciones.

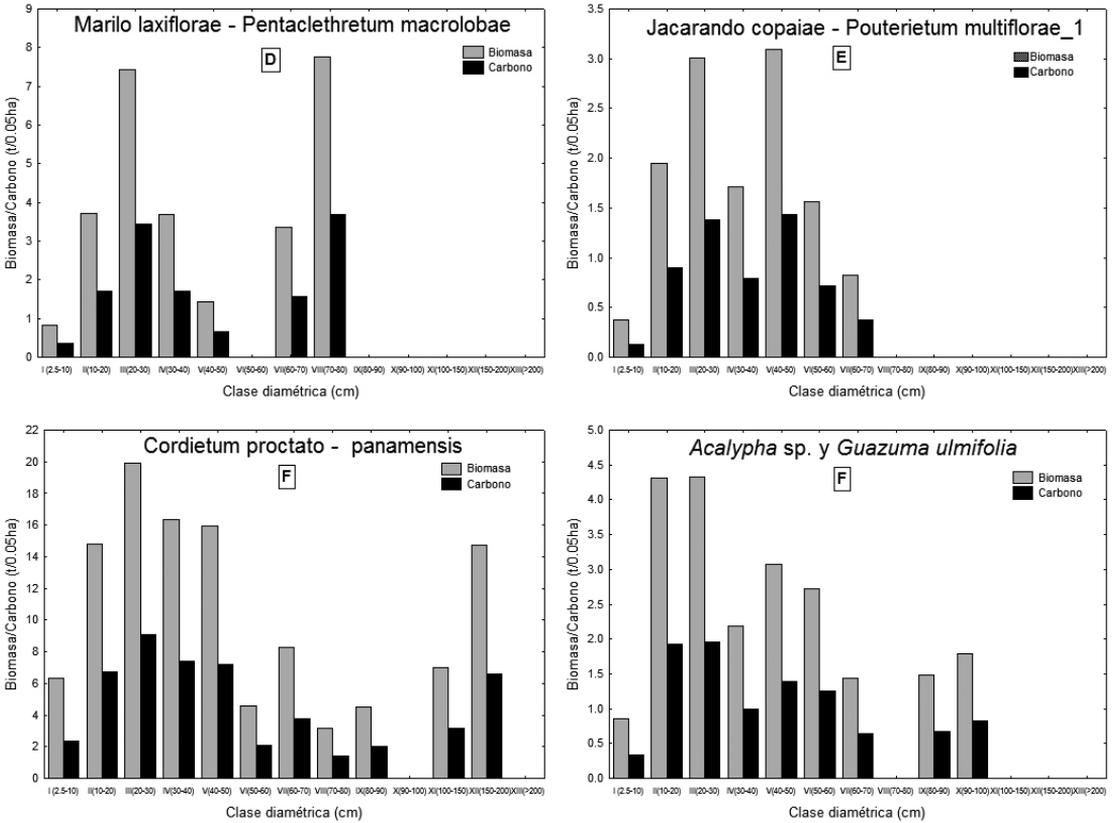

**Continuación Figura 378.** Distribución de la biomasa aérea por clase diamétrica de los individuos con DAP > a 2.5 cm de en los diferentes tipos de bosque de localidades del Sur y Noroccidente de Córdoba, Colombia. Las letras en los cuadrados indican los grupos de asociaciones.

Los bosques del grupo F corresponden a la asociación Cordietum proctato- Cordietum panamensis y a la comunidad de *Acalypha* sp. y *Guazuma ulmifolia* y se caracterizan porque la biomasa se concentra principalmente en las clases menores. En la primera asociación, las clases III(20-30), IV(30-40), V(40-50) y II(10-20) presentan alto contenido de biomasa y carbono que corresponde al 17, 14, 13 y 12% y equivale a 6.09 t/0.05ha en promedio para esta asociación. En la segunda comunidad, la mayor concentración de biomasa y carbono se presenta en las clases II(10-20) y III(20-30) con 19.53 y 19.44%, que corresponde a 2.88 t/0.05ha. Esta tendencia se relaciona con el estado sucesional que presenta esta comunidad, pues corresponde a bosques secundarios en su mayoría.

**Distribución de la biomasa y el carbono según los estratos**

La distribución de la biomasa en los diferentes estratos del bosque varió entre los diferentes tipos de bosque. La variación está dada por el número de individuos presente en cada estrato y por el tamaño de los individuos dominantes en cada estrato. La mayor acumulación de biomasa y de carbono se encontró en los estratos arbóreos en la mayoría de los tipos de bosque. En cinco bosques se concentra en el estrato arbóreo superior y en siete bosques en el arbóreo inferior; en un bosque se reparte por igual entre los estratos arbóreo superior e inferior y tan sólo en un bosque se concentra en el subarbóreo. En la figura 379 se aprecia esta variación. Los bosques que presentan una tendencia similar en la distribución de la biomasa y carbono se agruparon y se representan con una letra (A-F). El grupo A está conformado por las asociaciones Prestoeo decurrentis-Trichillietum poeppigi, Macrolobio ischnocalycis-Peltogynetum purpurea, Trichilio hirtae- Schizolobietum parahibi_1 y Jacarando copaiae-Pouterietum multiflorae_2. Se caracterizan porque la biomasa y el carbono se concentran principalmente en el estrato arbóreo superior seguido por el estrato arbóreo inferior con una diferencia muy marcada. En las dos primeras asociaciones, la biomasa se distribuye en un 71.26 y 71.02% en el estrato superior con valores medios de biomasa de 29.24 y 29.23 t/0.05ha. En la asociación Trichilio hirtae- Schizolobietum parahibi_1 se observa una acumulación del 70.39% de la biomasa en este estrato, que corresponde a 32.41 t/0.05ha. Los bosques de Jacarando copaiae-Pouterietum multiflorae_2 contienen en el estrato superior un 24.69% de la biomasa (12.34 t/0.05ha).

El grupo B lo conforman los bosques de la asociación Protio aracouchini-Viroletum elongatae; la biomasa se distribuye en cantidades similares en los estratos arbóreos superior e inferior, con valores de 16.21 y 16.16 t/0.05ha que equivalen al 48.64 y 48.67% de la biomasa aérea total de estos bosques.

El grupo C está constituido por las asociaciones Tovomito weddellianae- Quercetum humboldtii; Cappari odoratissimatis-Cavanillesietum platanifoliae; Cariniano pyriformis-Pentaplarietum doroteae y Trichilio hirtae- Schizolobietum parahibi_2, la biomasa y el carbono se acumulan en mayor cantidad en el estrato arbóreo inferior seguido por el estrato arbóreo superior. En la primera asociación se encontró una distribución de la biomasa de 16.64 y 11 t/0.05ha en los estratos arbóreos superior e inferior respectivamente, que equivalen al 33.29 y 22 % de la biomasa aérea total de estos bosques. En los bosques de la asociación Cappari odoratissimatis-Cavanillesietum platanifoliae se observó una acumulación de 11.32 t/0.05ha en el estrato arbóreo inferior y 9.51 t/0.05ha en el superior, que corresponde al 33.97 y 28.53%. En la asociación Cariniano pyriformis-Pentaplarietum doroteae se presentaron 10.1 y 8.19 t/0.05ha en los estratos arbóreo inferior y superior, respetivamente (50.24 y 40.7%). En los bosques de las asociaciones Trichilio hirtae- Schizolobietum parahibi_2, la biomasa en el estrato arbóreo inferior fue de 7.61 t/0.05ha (58.30%) y 3.94 t/0.05ha en el superior (30.18%).

En el grupo D se reúnen las asociaciones Cordietum proctato – panamensis y Marilo laxiflorae - Pentaclethretum macrolobae que presentan una mayor acumulación de biomasa y carbono en el estrato arbóreo inferior con un 52.5% (5.52 t/0.05ha), con participación similar de los estratos arbóreo superior y subarbóreo 25.29 (2.73 t/0.05ha) y 19.8% (2.07 t/0.05ha).

Las asociaciones Jacarando copaiae - Pouterietum multiflorae_1 y Mayno grandifoliae - Astrocaryetum malybo que conforman el grupo E no tenían representantes en el estrato arbóreo superior y la mayor proporción de biomasa y carbono se encuentra en el estrato arbóreo inferior, seguido por el subarbóreo. En la primera asociación la biomasa almacenada en el estrato inferior fue de 8.66 t/0.05ha en promedio y 2.72 t/0.05ha en el subarbóreo, valores que corresponden al 69 y 21%. En los bosques de la asociación Mayno grandifoliae - Astrocaryetum malybo, la biomasa en el estrato arbóreo inferior se cuantificó en 10.09 t/0.05ha (73.49%) y 2.34 t/0.05ha (17.04%).

El grupo F constituido por los bosques de la comunidad de *Acalypha* sp. y *Guazuma ulmifolia* presenta la mayor cantidad de biomasa y carbono en el estrato subarbóreo, seguido por el arbóreo inferior. En el estrato subarbóreo se encontró un promedio de 4.64 t/0.05ha que equivale al 62.83%, mientras que en el arbóreo inferior la biomasa sumó 2.52 t/0.05ha (34.05%).

El estrato arbustivo tiene baja representación en todos los tipos de bosque, no obstante en los bosques de las asociaciones Mayno grandifoliae - Astrocaryetum malybo, Jacarando copaiae - Pouterietum multiflorae_1 y la comunidad de *Acalypha* sp. y *Guazuma ulmifolia* este estrato alcanza mayor participación en la acumulación de biomasa y carbono (9.46%, 9.03% y 3.06% respectivamente).

Estas tendencias están en relación con los grados de intervención a los que se han visto sometidos los bosques estudiados, lo cual se refleja en la distribución incompleta en clases diamétricas y estratos (clases y estratos con ausencia de individuos), en la distribución diferente a los patrones descritos para

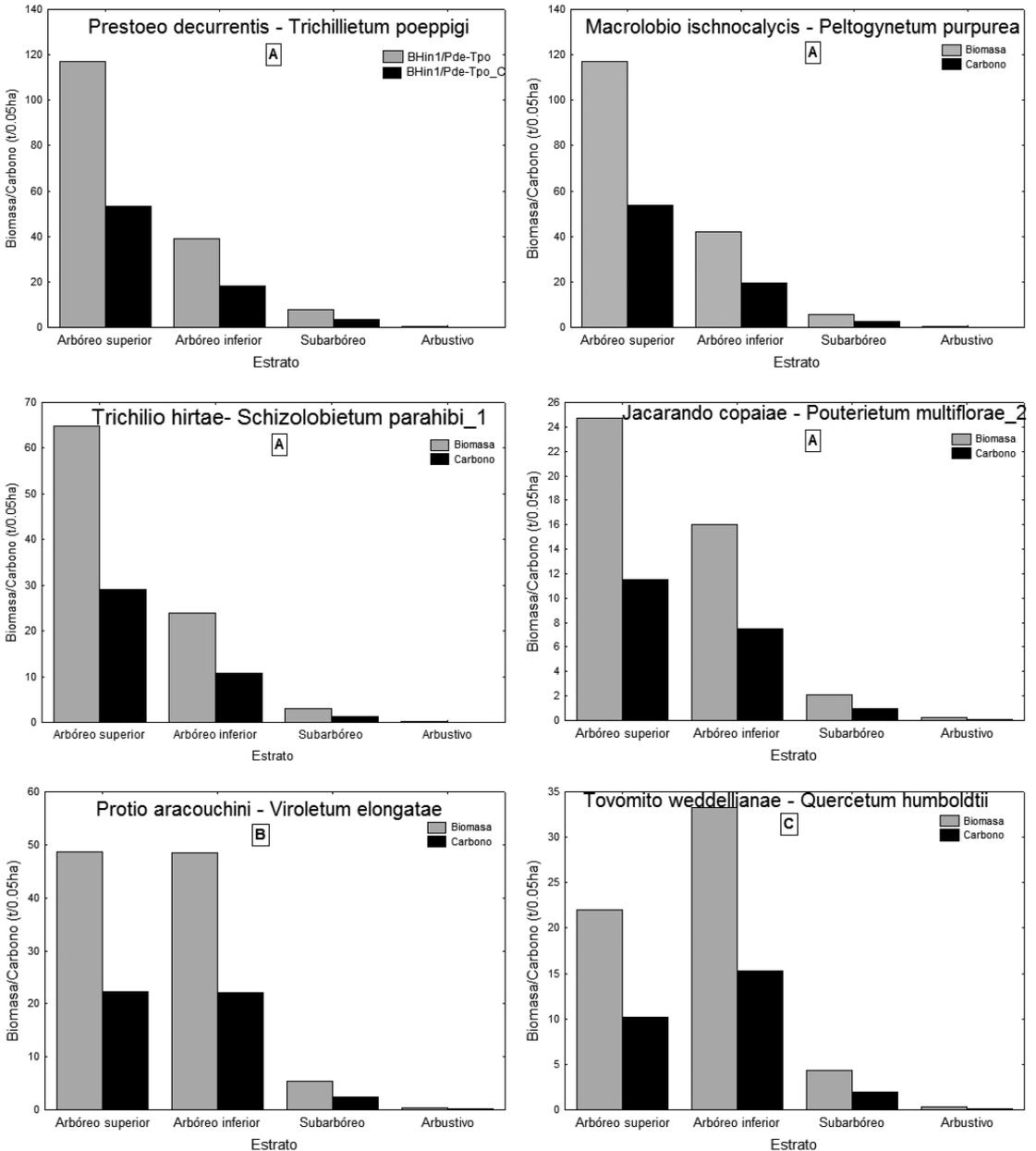

**Figura 379.** Distribución de la biomasa aérea por estrato para los individuos con DAP > a 2.5 cm en los diferentes tipos de bosque de localidades del Sur y Noroccidente de Córdoba, Colombia.

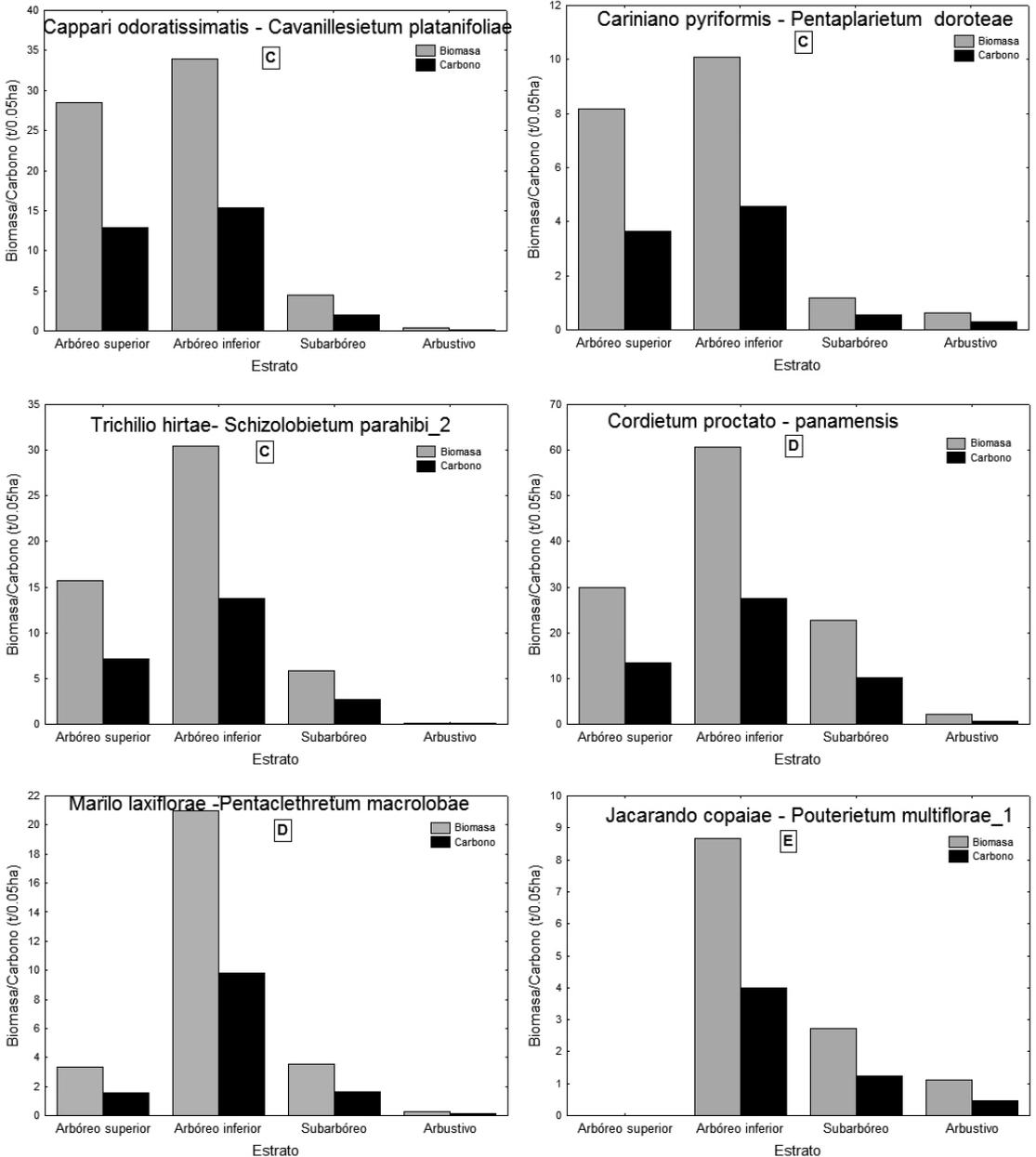

**Continuación Figura 379.** Distribución de la biomasa aérea por estrato para los individuos con DAP > a 2.5 cm en los diferentes tipos de bosque de localidades del Sur y Noroccidente de Córdoba, Colombia.

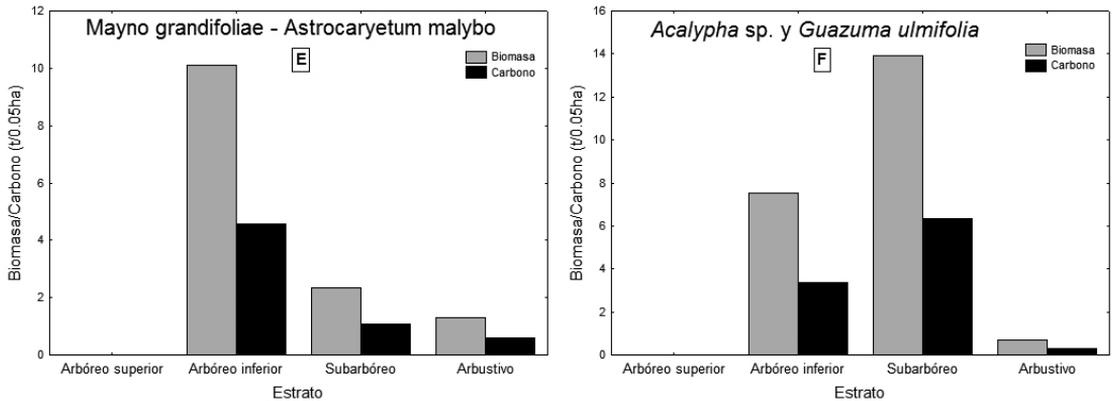

**Continuación Figura 379.** Distribución de la biomasa aérea por estrato para los individuos con DAP > a 2.5 cm en los diferentes tipos de bosque de localidades del Sur y Noroccidente de Córdoba, Colombia.

bosques sin intervención con tendencia a mayor acumulación de individuos en las clases menores (jota invertida) y en la observación en campo donde se aprecia la alteración con presencia de claros y árboles cortados.

La presencia de especies como *Cavanillesia platanifolia* que alcanza gran tamaño (diámetros superiores a 100 cm y alturas mayores a 20 m), conduce a la alta concentración de biomasa en los bosques de la asociación Trichilio hirtae- Schizolobietum parahibi_1, esta especie acumula el 34.74% de la biomasa en toda la asociación (Anexo 17) y el 74.37% en el estrato arbóreo superior (Tabla 171). Una tendencia similar muestra la especie *Peltogyne purpurea* en la asociación Macrolobio ischnocalycis - Peltogynetum purpurea que sola contribuye al 53.18 % de la biomasa aérea en estos bosques. En el anexo 17 se muestra la proporción de la biomasa almacenada en las especies dominantes en toda la asociación.

**Distribución de la biomasa en diferentes especies**

La biomasa y el carbono se distribuyen de manera desigual entre las especies en los diferentes tipos de bosque. La mayor acumulación se presentó en pocas especies. En el anexo 17 se muestran las cinco especies responsables de la mayor reserva de carbono a nivel general (sin distinguir entre estratos) para cada asociación, y su contribución (%) al total de las reservas. En los bosques de las asociaciones Prestoeo decurrentis - Trichillietum poeppigi, Protio aracouchini - Viroletum elongatae y Jacarando copaiae - Pouterietum multiflorae, la mayor contribución es dada por la especie *Dipteryx oleifera* 8.21 t/0.05ha (20.04%), 7.97 (23.25%) y 2.98t/0.05ha (16.16%) respectivamente. En la asociación Macrolobio ischnocalycis - Peltogynetum purpurea, la especie *Peltogyne purpurea* es la mayor contribuyente con 15.57 t/0.05ha (37.88%); en la asociación Tovomito weddellianae - Quercetum humboldtii es *Quercus humboldtii* la especie con mayor cantidad de biomasa 18.86 t/0.05ha (62.99%); en las asociaciones Cariniano pyriformis - Pentaplarietum doroteae y Cordietum proctato - Cordietum panamensis, la especie *Pentaplaris doroteae* es la mayor contribuyente a las reservas de biomasa y carbono 5.3 t/0.05ha (23.39%) y 2.31 t/0.05ha (22.14%) respectivamente. En la asociación Trichilio hirtae- Schizolobietum parahibi la mayor cantidad de biomasa esta almacenada por *Cavanillesia platanifolia* 8.36 t/0.05ha (34.74%); en la asociación Mayno grandifoliae - Astrocaryetum malybo, *Pseudobombax septenatum* es la especie con mayor reserva de biomasa 4.4 t/0.05ha (32.09%); en la asociación Marilo laxiflorae - Pentaclethretum macrolobae la especie con mayor reserva es *Pentaclethra macroloba* con 3.18 t/0.05ha (22.59%) y en la comunidad de *Acalypha* sp. *y Guazuma ulmifolia* la mayor cantidad de biomasa y carbono esta almacenada por *Guazuma ulmifolia* con 2.59 t/0.05ha (35.16%). En términos generales, entre el 40 y 80% de las reservas de biomasa y carbono son almacenadas por solo cinco especies en cada asociación (Anexo 17). En la tabla 171 se resumen las tres especies que contribuyen en mayor proporción a las reservas de biomasa y carbono en cada tipo de bosque, se distinguen los estratos.

**Tabla 171.** Especies dominantes según la biomasa en los diferentes tipos de bosque en localidades del Sur y Noroccidente de Córdoba.

| Alianza | Asociación | Estrato | Especie | Cobertura relativa | Densidad relativa | Area basal relativa | IPF relativo | Biomasa relativa | Carbono en la biomasa (%) |
|---|---|---|---|---|---|---|---|---|---|
| Eschweilero coriaceae - Pentacletrion macrolobae | Prestoeo decurrentis - Trichillietum poeppigi | Arbóreo superior | Dipteryx oleifera | 21.23 | 7.69 | 26.81 | 18.58 | 28.08 | 28.37 |
| | | | Anacardium excelsum | 10.62 | 7.69 | 25.81 | 14.71 | 17.45 | 17.22 |
| | | | Ceiba pentandra | 10.19 | 7.69 | 11.81 | 9.9 | 9.76 | 9.83 |
| | | Arbóreo inferior | Quararibea sp. 1 | 10.74 | 7.14 | 18.1 | 12 | 19.88 | 19.73 |
| | | | Sapotaceae sp. 3 | 1 | 1.79 | 12.5 | 5.1 | 13.36 | 13.25 |
| | | | Pentaclethra macroloba | 5.92 | 5.36 | 7.16 | 6.15 | 9.86 | 10.01 |
| | | | Dendropanax sp. 1 | 1.06 | 0.94 | 16.12 | 6.04 | 15.69 | 15.09 |
| | | Subarbóreo | Pentaclethra macroloba | 3.45 | 1.88 | 8.84 | 4.73 | 12.31 | 12.74 |
| | | | Anaxagorea crassipetala | 13.53 | 9.39 | 9.82 | 10.91 | 11.31 | 11.41 |
| | | | Eschweilera sp. 1 | 1.1 | 2.75 | 30.49 | 11.45 | 38.29 | 45.68 |
| | | Arbustivo | Tabernaemontana cymosa | 1.32 | 0.92 | 12.96 | 5.07 | 11.46 | 13.82 |
| | | | Arecaceae sp. 3 | 4.2 | 3.67 | 6.14 | 4.67 | 5.39 | 6.49 |
| | Macrolobio ischnocalycis - Peltogynetum purpurea | Arbóreo superior | Peltogyne purpurea | 44.15 | 55.56 | 46.55 | 48.75 | 53.18 | 52.99 |
| | | | Couratari guianensis | 10.9 | 11.11 | 20.48 | 14.16 | 14.29 | 14.46 |
| | | | Andira inermis | 20.43 | 11.11 | 15.76 | 15.77 | 13.45 | 13.43 |
| | | Arbóreo inferior | Eschweilera coriacea | 11.6 | 6.52 | 9.77 | 9.3 | 14.18 | 14.72 |
| | | | Licania sp.2 | 0.52 | 1.09 | 9.8 | 3.8 | 13.66 | 13.51 |
| | | | Macrolobium colombianum | 4.2 | 5.43 | 10.27 | 6.63 | 11 | 10.91 |
| | | | Macrolobium ischnocalyx | 21.11 | 17.56 | 23.68 | 20.78 | 24.59 | 24.75 |
| | | Subarbóreo | Helianthostylis sprucei | 2.3 | 0.76 | 4.01 | 2.36 | 4.31 | 4.41 |
| | | | Eschweilera coriacea | 3.17 | 3.05 | 2.85 | 3.02 | 3.71 | 3.89 |
| | | | Amphirrhox longifolia | 27.82 | 23.71 | 51.28 | 34.27 | 54.35 | 70.59 |
| | | Arbustivo | Gustavia sp.1 | 1.86 | 3.09 | 4.32 | 3.09 | 5.01 | 5.74 |
| | | | Macrolobium ischnocalyx | 8.25 | 6.19 | 3.94 | 6.13 | 3.88 | 2.69 |
| Billio roseae-Quercion humboldtii | Tovomito weddellianae - Quercetum humboldtii | Arbóreo superior | Quercus humboldtii | 75.97 | 66.67 | 82.23 | 74.96 | 85.74 | 85.65 |
| | | | Aspidosperma sp.3 | 9.89 | 16.67 | 10.55 | 12.37 | 9.02 | 9.02 |
| | | | Virola elongata | 14.13 | 16.67 | 7.21 | 12.67 | 5.24 | 5.32 |
| | | Arbóreo inferior | Quercus humboldtii | 33.56 | 20 | 47.23 | 33.6 | 56.53 | 56.45 |
| | | | Nectandra martinicensis | 3.51 | 2.86 | 10.66 | 5.67 | 9.24 | 9.23 |
| | | | Dendrobangia boliviana | 15.03 | 14.29 | 9.58 | 12.96 | 6.32 | 6.28 |
| | | | Tovomita weddelliana | 11.48 | 5.66 | 14.08 | 10.41 | 16.62 | 16.72 |
| | | Subarbóreo | Myrciaria floribunda | 11.3 | 9.43 | 9.29 | 10.01 | 10.89 | 10.9 |
| | | | Swartzia brachryrhachis | 2.8 | 3.77 | 6.58 | 4.39 | 9.21 | 8.91 |
| | | | Micropholis crotonoides | 7.09 | 3.88 | 7.88 | 6.28 | 8.91 | 11.34 |
| | | Arbustivo | Tovomita weddelliana | 6.34 | 3.88 | 5.9 | 5.38 | 7.96 | 10.98 |
| | | | Dendropanax arboreus | 6.34 | 6.8 | 7.99 | 7.04 | 7.53 | 8.45 |
| Brosimo utilis - Pentaclethrion macrolobae | Protio aracouchini - Viroletum elongatae | Arbóreo superior | Dipteryx oleifera | 20 | 14.29 | 19.12 | 17.8 | 30.58 | 31.58 |
| | | | Virola reidii | 41 | 28.57 | 37.77 | 35.78 | 27.25 | 25.82 |
| | | | Pouteria sp.5 | 10.67 | 14.29 | 16 | 13.65 | 13.22 | 13 |
| | | Arbóreo inferior | Brosimum guianense | 8.58 | 5.17 | 14.52 | 9.42 | 20.57 | 19.51 |
| | | | Dipteryx oleifera | 9.65 | 3.45 | 12.41 | 8.5 | 18.32 | 18.89 |
| | | | Jacaranda copaia | 5.15 | 3.45 | 6.94 | 5.18 | 6.26 | 6.78 |
| | | | Amphirrhox longifolia | 14.82 | 19.55 | 11.62 | 15.33 | 9.7 | 9.75 |
| | | Subarbóreo | Helicostylis tomentosa | 3.3 | 3.76 | 7.39 | 4.82 | 6.5 | 6.41 |
| | | | Caryocar amygdaliferum | 0.77 | 1.5 | 6.13 | 2.8 | 5.77 | 5.79 |
| | | | Euphorbiaceae sp.1 | 1.73 | 1.55 | 29.77 | 11.01 | 29.08 | 45.65 |
| | | Arbustivo | Amphirrhox longifolia | 16.59 | 7.75 | 14.23 | 12.86 | 14.51 | 19.13 |
| | | | Protium aracouchini | 6.22 | 5.43 | 3.67 | 5.11 | 3.54 | 2.31 |
| Brosimo utilis - Pentaclethrion macrolobae | Jacarando copaiae - Pouterietum multiflorae_1 | Arbóreo inferior | Caesalpiniaceae sp.2 | 4.96 | 6.25 | 20.16 | 10.46 | 18.06 | 18.05 |
| | | | Pouteria buenaventurensis | 17.36 | 12.5 | 18.44 | 16.1 | 16.33 | 16.2 |
| | | | Ocotea cernua | 13.22 | 6.25 | 13.4 | 10.96 | 15.07 | 15.06 |
| | | Subarbóreo | Senefeldera testiculata | 6.68 | 6 | 9.42 | 7.37 | 12.32 | 12.37 |
| | | | Casearia arguta | 6.68 | 6 | 9.35 | 7.34 | 10.16 | 10.2 |
| | | | Fabaceae sp.2 | 8.24 | 4 | 7.89 | 6.71 | 8.51 | 8.55 |

**Continuación Tabla 171.** Especies dominantes según la biomasa en los diferentes tipos de bosque en localidades del Sur y Noroccidente de Córdoba.

| Alianza | Asociación | Estrato | Especie | Cobertura relativa | Densidad relativa | Area basal relativa | IPF relativo | Biomasa relativa | Carbono en la biomasa (%) |
|---|---|---|---|---|---|---|---|---|---|
| Brosimo utilis - Pentacletrion macrolobae | Jacarando copaiae - Pouterietum multiflorae_1 | Arbustivo | *Tovomita stylosa* | 3.87 | 2.63 | 59.5 | 22 | 72.47 | 78.12 |
| | | | *Matisia bracteolosa* | 1.93 | 1.32 | 17.3 | 6.85 | 10.02 | 10.45 |
| | | | *Dendrobangia boliviana* | 1.93 | 1.32 | 3.14 | 2.13 | 2.63 | 2.82 |
| | Jacarando copaiae - Pouterietum multiflorae_2 | Arbóreo superior | *Dipteryx oleifera* | 12.62 | 12.5 | 28.29 | 17.8 | 33.95 | 34.49 |
| | | | *Dialium guianense* | 25.49 | 12.5 | 23.76 | 20.58 | 29.65 | 29.7 |
| | | | *Pouteria multiflora* | 15.29 | 12.5 | 20.65 | 16.15 | 12.53 | 12.21 |
| | | Arbóreo inferior | *Eschweilera pittieri* | 9.34 | 10.26 | 9.86 | 9.82 | 14.38 | 14.81 |
| | | | *Pentaclethra macroloba* | 2.64 | 2.56 | 8.44 | 4.55 | 12.99 | 13.19 |
| | | | *Dendrobangia boliviana* | 15.37 | 7.69 | 10.71 | 11.26 | 10.42 | 10.2 |
| | | | *Iryanthera hostmannii* | 9.67 | 9.09 | 12.79 | 10.51 | 16.17 | 16.19 |
| | | Subarbóreo | *Inga* sp.2 | 4.36 | 2.6 | 7.2 | 4.72 | 9.03 | 9.65 |
| | | | *Dipteryx oleifera* | 2.72 | 1.3 | 6.44 | 3.49 | 8.83 | 9.13 |
| | | | *Wettinia hirsuta* | 10.72 | 11.34 | 15.73 | 12.59 | 24.17 | 37.53 |
| | | Arbustivo | *Guarea kunthiana* | 7.61 | 8.25 | 14.27 | 10.04 | 10.16 | 13.7 |
| | | | *Faramea* sp.2 | 5.19 | 5.15 | 6.19 | 5.51 | 6.1 | 5.69 |
| | Marilo laxiflorae - Pentaclethretum macrolobae | Arbóreo superior | *Brosimum utile* | 10 | 1 | 0.3 | 0 | 3.35 | 1.57 |
| | | Arbóreo inferior | *Pentaclethra macroloba* | 26.49 | 17.78 | 29.55 | 24.61 | 29.18 | 29.49 |
| | | | *Sloanea tuerckheimii* | 2.49 | 2.22 | 16.27 | 7 | 19.78 | 20 |
| | | | *Sloanea zuliaensis* | 6.18 | 4.44 | 6.33 | 5.65 | 9.28 | 9.35 |
| | | | *Iryanthera hostmannii* | 21.65 | 10 | 14.6 | 15.42 | 14.35 | 14.63 |
| | | Subarbóreo | *Melastomataceae* sp.2 | 2.01 | 2.22 | 5.38 | 3.2 | 6.75 | 6.75 |
| | | | *Pentaclethra macroloba* | 4.3 | 5.56 | 7.14 | 5.67 | 6.24 | 6.4 |
| | | | *Casearia arguta* | 2.9 | 1.72 | 18.58 | 7.74 | 22.53 | 27.3 |
| | | Arbustivo | *Amphirrhox longifolia* | 16.88 | 12.07 | 12.19 | 13.71 | 12.21 | 12.8 |
| | | | *Dendropanax arboreus* | 4.36 | 3.45 | 6.81 | 4.87 | 6.36 | 7.23 |
| Astrocaryo malybo - Cavanillesion platanifoliae | Cariniano pyriformis - Pentaplarietum doroteae | Arbóreo superior | *Cavanillesia platanifolia* | 0 | 50 | 62.56 | 37.52 | 59.23 | 59.64 |
| | | | *Pentaplaris doroteae* | 100 | 50 | 37.44 | 62.48 | 40.77 | 40.36 |
| | | Arbóreo inferior | *Castilla elastica* | 24 | 13.33 | 34.02 | 23.78 | 30.72 | 31.28 |
| | | | *Pentaplaris doroteae* | 13.71 | 6.67 | 17.35 | 12.58 | 19.43 | 18.85 |
| | | | *Croton* sp.1 | 10 | 6.67 | 15.71 | 10.79 | 17.68 | 17.93 |
| | | | *Melastomataceae* sp.3 | 26.21 | 12 | 25.25 | 21.15 | 27.08 | 26.99 |
| | | Subarbóreo | *Mouriri completens* | 6.45 | 4 | 10.56 | 7 | 12.9 | 12.85 |
| | | | *Annonaceae* sp.2 | 8.06 | 4 | 10.24 | 7.44 | 12.51 | 12.47 |
| | | | *Astrocaryum malybo* | 87.84 | 69.62 | 82 | 79.82 | 80.69 | 82.72 |
| | | Arbustivo | *Faramea capillipes* | 6.22 | 16.46 | 11.74 | 11.47 | 12.87 | 12.82 |
| | | | *Hirtella americana* | 0.16 | 1.27 | 0.92 | 0.78 | 1.59 | 1.65 |
| | Mayno grandifoliae - Astrocaryetum malybo | Arbóreo inferior | *Pseudobombax septenatum* | 27.39 | 15 | 44.42 | 28.93 | 42.98 | 41.9 |
| | | | *Schizolobium parahyba* | 7.26 | 5 | 12.54 | 8.27 | 13.33 | 13.57 |
| | | | *Bursera simaruba* | 15.77 | 20 | 10.85 | 15.54 | 8.5 | 8.61 |
| | | | *Cupania* sp.1 | 3.43 | 2 | 10.89 | 5.44 | 17.63 | 17.72 |
| | | Subarbóreo | *Cordia* sp.2 | 8.24 | 8 | 10.6 | 8.94 | 12.38 | 12.33 |
| | | | *Annona* sp.1 | 13.96 | 6 | 11.79 | 10.58 | 10.55 | 10.43 |
| | | | *Astrocaryum malybo* | 73.88 | 74.71 | 88.75 | 79.11 | 86.57 | 87.4 |
| | | Arbustivo | *Gustavia superba* | 17 | 12.64 | 7.67 | 12.44 | 8.93 | 8.64 |
| | | | *Bactris pilosa* | 0.25 | 1.15 | 1.2 | 0.87 | 1.5 | 1.53 |
| Alianza no definida | Trichilio hirtae - Schizolobietum parahibi_1 | Arbóreo superior | *Cavanillesia platanifolia* | 31.91 | 25 | 61.8 | 39.57 | 74.37 | 74.74 |
| | | | *Pseudobombax septenatum* | 25.53 | 25 | 21.12 | 23.88 | 11.87 | 11.59 |
| | | | *Pentaplaris doroteae* | 21.28 | 25 | 11.8 | 19.36 | 9.88 | 9.7 |
| | | Arbóreo inferior | *Pentaplaris doroteae* | 32.36 | 33.33 | 31.52 | 32.4 | 32.38 | 31.77 |
| | | | *Machaerium goudotii* | 4.97 | 3.03 | 11.3 | 6.43 | 13.72 | 13.99 |
| | | | *Vitex* sp.1 | 8.94 | 3.03 | 9.81 | 7.26 | 10.82 | 10.56 |
| | | | *Anacardium excelsum* | 15.29 | 1.45 | 38.36 | 18.37 | 32.86 | 33.17 |
| | | Subarbóreo | *Zanthoxylum setulosum* | 6.21 | 2.9 | 11.25 | 6.79 | 13.27 | 13.43 |
| | | | *Pouteria subrotata* | 17.52 | 15.94 | 9.46 | 14.31 | 11.59 | 11.46 |

**Continuación Tabla 171.** Especies dominantes según la biomasa en los diferentes tipos de bosque en localidades del Sur y Noroccidente de Córdoba.

| Alianza | Asociación | Estrato | Especie | Cobertura relativa | Densidad relativa | Area basal relativa | IPF relativo | Biomasa relativa | Carbono en la biomasa (%) |
|---|---|---|---|---|---|---|---|---|---|
| Alianza no definida | Trichilio hirtae-Schizolobietum parahibi_1 | Arbustivo | *Myrtaceae* sp.9 | 16.3 | 7.59 | 10.94 | 11.61 | 16.5 | 21.33 |
| | | | *Faramea parvibracteata* | 21.6 | 21.52 | 17.2 | 20.11 | 16.04 | 9.16 |
| | | | *Pouteria subrotata* | 6.52 | 3.8 | 9.23 | 6.52 | 11.3 | 15.85 |
| Alianza no definida | Trichilio hirtae-Schizolobietum parahibi_2 | Arbóreo superior | *Trichilia hirta* | 79.77 | 66.67 | 63.34 | 69.93 | 75.29 | 75.32 |
| | | | *Bursera simaruba* | 20.23 | 33.33 | 36.66 | 30.07 | 24.71 | 24.68 |
| | | Arbóreo inferior | *Trichilia hirta* | 10.82 | 8.16 | 16.48 | 11.82 | 22.4 | 22.46 |
| | | | *Sterculia* sp.2 | 10.63 | 14.29 | 16.7 | 13.87 | 14.25 | 14.05 |
| | | | *Fabaceae* sp.4 | 3.64 | 4.08 | 5.04 | 4.25 | 7.35 | 7.37 |
| | | | *Bombacaceae* sp.1 | 2.11 | 2.3 | 10.05 | 4.82 | 12.61 | 12.61 |
| | | Subarbóreo | *Schizolobium* sp.1 | 20.28 | 8.05 | 12.79 | 13.7 | 12.21 | 12.29 |
| | | | *Trichilia acuminata* | 11.39 | 13.79 | 8.29 | 11.16 | 10.22 | 10.39 |
| | | | *Trichilia hirta* | 10.63 | 5.41 | 18.23 | 11.42 | 20.05 | 24.85 |
| | | Arbustivo | *Faramea occidentalis* | 23.78 | 24.32 | 19.39 | 22.5 | 19.41 | 18.4 |
| | | | *Ampelocera macphersonii* | 18.24 | 18.92 | 13.49 | 16.88 | 18.07 | 16.86 |
| Cratevo tapiae - Cavanillesion platanifoliae | Cordietum proctato - Cordietum panamensis | Arbóreo superior | *Pentaplaris doroteae* | 31.25 | 16.67 | 55.46 | 34.46 | 49.26 | 48.95 |
| | | | *Cavanillesia platanifolia* | 54.69 | 50 | 42 | 48.9 | 46.76 | 46.99 |
| | | | *Pterocarpus acapulcensis* | 7.81 | 16.67 | 2.52 | 9 | 3.98 | 4.06 |
| | | Arbóreo inferior | *Pentaplaris doroteae* | 11.08 | 9.8 | 13.53 | 11.47 | 14.11 | 13.65 |
| | | | *Bursera simaruba* | 9.01 | 8.5 | 12.11 | 9.87 | 8.7 | 8.69 |
| | | | *Acanthaceae* sp.2 | 9.52 | 7.19 | 6.71 | 7.81 | 7.69 | 7.72 |
| | | | *Pentaplaris doroteae* | 1.2 | 1.05 | 8.12 | 3.46 | 8.49 | 8.32 |
| | | Subarbóreo | *Cordia panamensis* | 4.81 | 4.06 | 7.5 | 5.46 | 7.66 | 7.82 |
| | | | *Alseis blackiana* | 4.98 | 4.36 | 6.5 | 5.28 | 6.52 | 6.59 |
| | | | *Guettarda acreana* | 3.33 | 1.83 | 3.9 | 3.02 | 7.3 | 9.93 |
| | | Arbustivo | *Bignoniaceae* sp.2 | 1.05 | 0.46 | 8.89 | 3.47 | 7.14 | 10.38 |
| | | | *Rosenbergiodendron formosum* | 2.38 | 1.68 | 5.35 | 3.14 | 6.03 | 8.2 |
| | Cappari odoratissimatis - Cavanillesietum platanifoliae | Arbóreo superior | *Bombacaceae* sp.1 | 35.71 | 25 | 46.02 | 35.58 | 50.21 | 50.29 |
| | | | *Cavanillesia platanifolia* | 53.57 | 50 | 51.15 | 51.57 | 46.36 | 46.24 |
| | | | *Pterocarpus acapulcensis* | 10.71 | 25 | 2.83 | 12.85 | 3.43 | 3.48 |
| | | Arbóreo inferior | *Ruprechtia costata* | 28.74 | 28.85 | 23.19 | 26.92 | 27.45 | 27.6 |
| | | | *Bombacaceae* sp.1 | 6.49 | 3.85 | 22.77 | 11.03 | 19.19 | 19.25 |
| | | | *Cavanillesia platanifolia* | 3.48 | 1.92 | 14.35 | 6.58 | 12.35 | 12.31 |
| | | | *Machaerium milleflorum* | 19.74 | 9.68 | 11.54 | 13.65 | 14.94 | 14.99 |
| | | Subarbóreo | *Alseis blackiana* | 5.4 | 7.26 | 8.88 | 7.18 | 8.71 | 8.81 |
| | | | *Ampelocera macphersonii* | 6.84 | 8.06 | 7.29 | 7.4 | 8.38 | 8.16 |
| | | | *Trichilia acuminata* | 11.77 | 10.92 | 16.54 | 13.08 | 22.08 | 28.37 |
| | | Arbustivo | *Psychotria microdon* | 7.77 | 11.76 | 12.15 | 10.56 | 12.81 | 14.01 |
| | | | *Crateva tapia* | 4.22 | 2.52 | 18.91 | 8.55 | 11.09 | 14.55 |
| Alianza no definida | *Acalypha* sp. y *Guazuma ulmifolia* | Arbóreo inferior | *Tectona grandis* | 21.14 | 32.35 | 24.13 | 25.88 | 14.25 | 13.33 |
| | | | *Tabebuia rosea* | 22.87 | 23.53 | 18.49 | 21.63 | 22.39 | 23.58 |
| | | | *Spondias mombin* | 11.34 | 5.88 | 20.38 | 12.53 | 21.37 | 21.68 |
| | | | *Guazuma ulmifolia* | 25.88 | 21.82 | 61.25 | 36.31 | 54.96 | 55.3 |
| | | Subarbóreo | *Caesalpinia glabrata* | 9.52 | 9.09 | 2.92 | 7.18 | 6.29 | 6.1 |
| | | | *Tabebuia rosea* | 6.02 | 8.18 | 4.19 | 6.13 | 5.1 | 5.26 |
| | | | *Randia armata* | 18.19 | 9.89 | 28.21 | 18.76 | 35.4 | 38.71 |
| | | Arbustivo | *Guazuma ulmifolia* | 7.68 | 10.99 | 17.72 | 12.13 | 16.18 | 16.82 |
| | | | *Myrospermum frutescens* | 3.54 | 2.2 | 10.23 | 5.32 | 7.75 | 8.59 |

Estructura, biomasa aérea y carbono almacenado en los bosques de CórdobaLos resultados encontrados en este estudio se asemejan a los mencionados por Kirby & Potvin (2007), en un estudio sobre la variación del almacenamiento de carbono entre especies en Iperi Embera (Panamá) el cual muestra que la contribución relativa de las especies al almacenamiento de carbono por hectárea en bosques y sistemas agroforestales, es altamente sesgada y frecuentemente no es proporcional a la abundancia de especies.

**Distribución de la biomasa en familias**

Al igual que en las especies, la biomasa también se acumula en mayor proporción en pocas familias. En la tabla 172 se muestra esta tendencia, casi un 80% de la biomasa se concentra principalmente en 20 familias en los bosques estudiados.

**Biomasa y carbono en el gradiente de precipitación**

La biomasa con relación al gradiente de precipitación presentó una variación entre 34.18 t/0.05ha para los bosques en clima súper húmedo hasta 7.39 t/0.05ha para los semihúmedos. Los bosques en clima muy húmedo y húmedo presentaron valores similares de 21.15 y 21.39 t/0.05ha (Tabla 173). La prueba de Kruskal-Wallis encontró diferencias significativas de la biomasa entre los climas (H=19.38; p=0.0002) (Figura 380).

Esta tendencia se relaciona a su vez con el grado de disturbio al que han sido sometidos los bosques en estudio, dado que en cada clima hay diferentes grados de intervención. La intervención es más pronunciada a medida que se avanza de muy

**Tabla 172.** Familias dominantes según la proporción de biomasa en los diferentes tipos de bosque en localidades del Sur y Noroccidente de Córdoba.
Las familias con mayor biomasa en cada asociación aparecen resaltadas.

| Familia | Asociación/comunidad | | | | | | | | | | | |
|---|---|---|---|---|---|---|---|---|---|---|---|---|
| | BHin1/ Pde-Tpo | BHri1/ Mis-Ppu | BHtf2/ Jco-Pmu | BHtf1/ Twe-Qhu | Bmhtf1/ Par-Vel | Bmhtf2/ Mla-Pma | Bmhtf2/ Cpy-Pdo | Bmhtf2/ Mgr-Ama | Bhtf1/ Thi-Spa | Bhtf2/ Cod-Cpl | Bhtf2/ Cpr-Cpa | Bttf3/ Asp-Gul |
| Anacardiaceae | 16.6 | 0 | 0 | 0 | 0 | 0 | 2.74 | 5.1 | 1.09 | 1.68 | 0.78 | 8.16 |
| Apocynaceae | 0.48 | 0.4 | 0.02 | 6.4 | 1.11 | 2.2 | 0 | 0 | 0.51 | 0 | 0.22 | 0.27 |
| Arecaceae | 0.39 | 0.11 | 0.74 | 0 | 0.32 | 0.99 | 2.74 | 8.81 | 0.68 | 0 | 0.16 | 0.45 |
| Bignoniaceae | 2.17 | 0 | 3.06 | 0 | 3.1 | 0.07 | 3.53 | 0 | 0 | 0 | 4.82 | 11.01 |
| Bombacaceae | 6.95 | 9.34 | 0.2 | 0 | 0.82 | 4.76 | 24.13 | 32.48 | 66.29 | 56.92 | 13.47 | 3.47 |
| Caesalpiniaceae | 0 | 5.22 | 16.62 | 0 | 5.7 | 0.04 | 1.14 | 12.67 | 1.55 | 0.21 | 1.74 | 4.74 |
| Chrysobalanaceae | 0.21 | 8.02 | 0.32 | 0.7 | 0 | 0.39 | 0.05 | 0 | 0 | 0 | 0 | 0 |
| Elaeocarpaceae | 0.3 | 0.86 | 0 | 0.1 | 5.02 | 14.76 | 0 | 0 | 0 | 0 | 0 | 0 |
| Fabaceae | 22.81 | 48.56 | 18.34 | 0.67 | 25.58 | 0.39 | 0 | 0 | 1.43 | 6.03 | 6.36 | 0.95 |
| Fagaceae | 0 | 0 | 0 | 62.87 | 0 | 0 | 0 | 0 | 0 | 0 | 0 | 0 |
| Lauraceae | 0.01 | 0.02 | 5.52 | 5.7 | 0.82 | 0.78 | 0 | 0.58 | 0.09 | 0 | 0 | 0 |
| Lecythidaceae | 1.97 | 14.16 | 5.54 | 0.42 | 3.82 | 2.87 | 1.24 | 2.55 | 0.07 | 0.01 | 0.69 | 3.92 |
| Meliaceae | 1.84 | 0.14 | 2.84 | 1.73 | 0.25 | 0.6 | 0 | 0 | 19.68 | 0.64 | 1.97 | 3.92 |
| Mimosaceae | 3.6 | 2.47 | 5.23 | 0.65 | 2.16 | 25.02 | 0.45 | 0.07 | 1.89 | 0.31 | 3.64 | 7.13 |
| Moraceae | 0.97 | 0.95 | 3.34 | 0 | 12.59 | 14.94 | 15.62 | 0.15 | 0.63 | 0 | 2.86 | 1.17 |
| Myristicaceae | 1.3 | 0.63 | 3.18 | 4.22 | 19.82 | 5.11 | 0 | 1.31 | 0 | 0 | 0 | 0 |
| Polygonaceae | 6.59 | 0 | 0.14 | 0 | 0 | 0 | 0 | 0 | 0.43 | 15.25 | 4.38 | 0 |
| Sapotaceae | 8.68 | 1.07 | 11.11 | 1.98 | 7.21 | 1.63 | 0 | 2.4 | 0.96 | 3.89 | 2.69 | 0 |
| Sterculiaceae | 0.44 | 0.18 | 1.78 | 0 | 1.13 | 0 | 0 | 0 | 1.92 | 0.01 | 2.63 | 35.18 |
| Tiliaceae | 5.71 | 0.01 | 0 | 0 | 0.5 | 0 | 26.37 | 0 | 14.24 | 2.36 | 21.93 | 0.09 |
| Contribución total (%) | 81.02 | 92.14 | 77.98 | 85.44 | 89.95 | 74.55 | 78.01 | 66.12 | 111.46 | 87.31 | 68.34 | 80.46 |

Convenciones

| Código | Asociación | Código | Asociación |
|---|---|---|---|
| BHin1/Pde-Tpo | Prestoeo decurrentis-Trichillietum poeppigi | Bmhtf2/Cpy-Pdo | Cariniano pyriformis-Pentaplarietum doroteae |
| BHri1/Mis-Ppu | Macrolobio ischnocalycis-Peltogynetum purpurea | Bmhtf2/Mgr-Ama | Mayno grandifoliae-Astrocaryetum malybo |
| BHtf2/Jco-Pmu | Jacarando copaiae-Pouterietum multiflorae | Bhtf1/Thi-Spa | Trichilio hirtae-Schizolobietum parahibi |
| BHtf1/Twe-Qhu | Tovomito weddellianae-Quercetum humboldtii | Bhtf2/Cod-Cpl | Cappari odoratissimatis-Cavanillesietum platanifoliae |
| Bmhtf1/Par-Vel | Protio aracouchini-Viroletum elongatae | Bhtf2/Cpr-Cpa | Cordietum proctato- panamensis |
| Bmhtf2/Mla-Pma | Marilo laxiflorae-Pentaclethretum macrolobae | Bttf3/Asp-Gul | *Acalypha sp. y* Guazuma ulmifolia |

954

húmedo, húmedo a semihúmedo, que explica la alta variabilidad observada en los bosques de estos ambientes.

Los bosques en climas super húmedos presentan grados de intervención de nulo a alto, predomina el grado bajo. En el clima muy húmedo, el grado de intervención es de bajo a medio. En el clima húmedo de nulo a alto, predomina el medio. En el clima semi húmedo solo se muestrearon sitios alterados (grado alto de intervención), que se relaciona con la baja cantidad de biomasa.

**Tabla 173.** Valores medios de biomasa y de carbono de los bosques por clima en el gradiente de precipitación súper húmedo-semihúmedo en localidades de Córdoba.

| Clima | Biomasa (t/0.05m2) | Carbono (t/0.05m2) |
|---|---|---|
| Súper húmedo | 34.18 | 15.68 |
| Muy húmedo | 21.15 | 9.62 |
| Húmedo | 21.39 | 9.64 |
| Semihúmedo | 7.39 | 3.34 |

## CONSIDERACIONES FINALES

Se observó alta variación en los aspectos de la estructura y por consiguiente, en las reservas de biomasa y carbono entre los diferentes tipos de bosque del departamento de Córdoba. Este comportamiento está relacionado con los procesos de sucesión natural, con los factores ambientales y con la presión antrópica.

La biomasa varió entre 7.39±1.79 y 41.15±21.81 t/0.05ha entre los diferentes tipos de bosque. Fue menor para el bosque de la comunidad de *Acalypha* sp. y *Guazuma ulmifolia* y mayor para los bosques de la asociación Macrolobio ischnocalycis - Peltogynetum purpurea. De acuerdo a la cantidad de biomasa y carbono almacenados, los bosques analizados pertenecientes a las diferentes asociaciones se agruparon en cinco categorías: en la clase *muy altas* (>36.79-44.14 t/0.05ha) se ubican los bosques de las asociaciones Prestoeo decurrentis - Trichillietum poeppigi, Macrolobio ischnocalycis - Peltogynetum purpurea presentaron las mayores reservas de biomasa y carbono 41.04±16.77 t/0.05ha y se ubican. En la categoría *altas* (>29.44-36.79 t/0.05ha) reservas de biomasa y de carbono se ubicaron los bosques de las asociaciones Protio aracouchini - Viroletum elongatae y Tovomito weddellianae - Quercetum humboldtii, con valores comprendidos entre 30.01±5.15 y 34.33±11.73 t/0.05ha. Los bosques de las asociaciones Trichilio hirtae- Schizolobietum parahibi y Cappari odoratissimatis - Cavanillesietum platanifoliae se agruparon en la categoría reservas de biomasa y carbono *medias* (>22.09-29.44 t/0.05ha) con valores entre 22.45±11.69 y 24.06±23.26 t/0.05ha. Los bosques de las asociaciones Cariniano pyriformis - Pentaplarietum doroteae y Jacarando copaiae - Pouterietum multiflorae se clasificaron bajo la denominación reservas *bajas* (>14.74-22.09 t/0.05ha) con valores de biomasa entre 18.54±7.94 y 20.1±2.05 t/0.05ha. Los bosques de las asociaciones Marilo laxiflorae - Pentaclethretum macrolobae, Mayno grandifoliae - Astrocaryetum malybo, Cordietum proctato - panamensis y la comunidad de *Acalypha* sp. y *Guazuma ulmifolia* presentan la menor biomasa; se agruparon en la denominación reservas de biomasa y carbono muy *bajas* (7.39-14.74 t/0.05ha) con valores que oscilan entre 7.39±1.79 y 14.09±2.25 t/0.05ha.

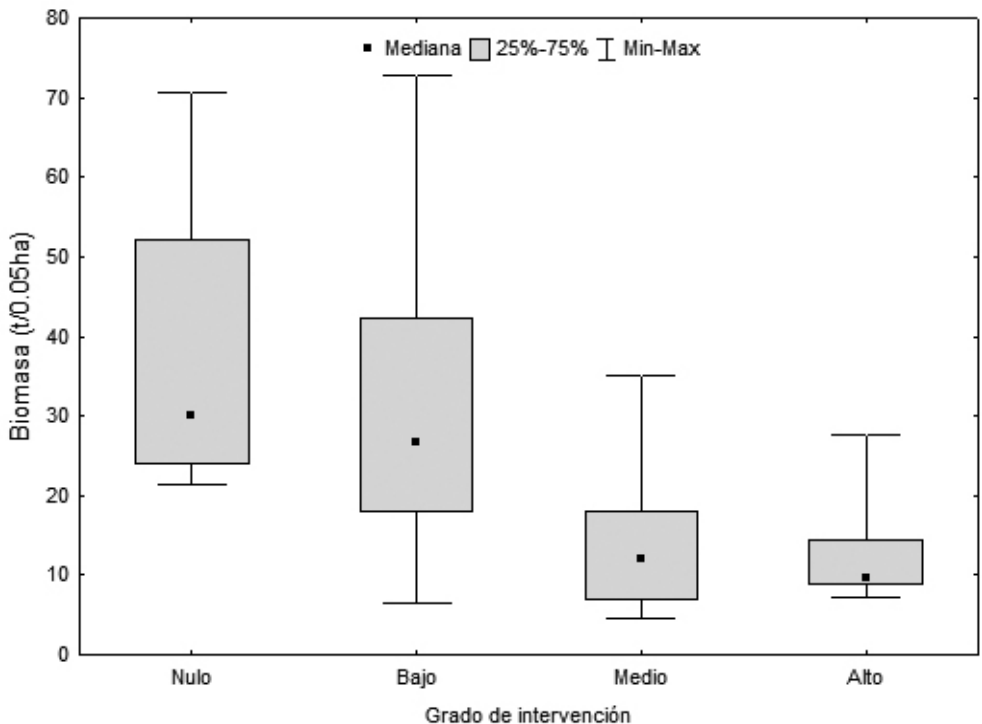

**Figura 380.** Variación de la biomasa con relación al gradiente de precipitación súper húmedo-semihúmedo en localidades de Córdoba.

Las clases diamétricas mayores (>70cm) y los estratos arbóreos (superior e inferior) concentran la mayor cantidad de biomasa y de carbono en la mayoría de los bosques evaluados. Las asociaciones Prestoeo decurrentis - Trichillietum poeppigi y Macrolobio ischnocalycis - Peltogynetum purpurea contienen la mayor reserva de biomasa y carbono en la clase diamétrica VI (>100cm) con 41.03%. Las asociaciones Protio aracouchini - Viroletum elongatae y Tovomito weddellianae - Quercetum humboldtii contienen las mayores reservas en la clase V (70-100cm). Esta distribución explica los valores altos de biomasa y carbono observados en estas asociaciones. En contraste, la asociación Cordietum proctato - panamensis y la comunidad de *Acalypha* sp. y *Guazuma ulmifolia* tienen mayor cantidad de biomasa acumulada (57.91% y 62.71% respectivamente) en las clases diamétricas menores –clases II (10-20cm) y III (30-50cm)-. Esta tendencia tiene por supuesto, estrecha relación con el grado de intervención de cada bosque.

Las diferencias entre la biomasa de los bosques analizados se debe principalmente a la presencia de los elementos árboreos del estrato superior. Con los resultados obtenidos, se podría concluir que la ausencia de elementos de gran porte en algunos levantamientos de bosques en recuperación es sustituída por la concentración de biomasa en estratos y clases diamétricas inferiores, como es el caso de la comunidad de *Acalypha* sp. y *Guazuma ulmifolia* y la asociación Cordietum proctato - panamensis*;* sin embargo, debido a los diferentes grados de transformación e intervención de estas localidades, se requieren más sitios de muestreo para comprobar esta afirmación.

En la distribución de la biomasa aérea y por ende en las reservas de carbono en dicha biomasa, se observa la dominancia de pocos individuos que concentran alta cantidad de carbono; sin embargo las clases de DAP y altura inferiores tienen un papel fundamental en el almacenamiento de carbono debido al alto número de individuos que contienen y a que están en proceso de crecimiento y algunos serán los individuos dominantes.

Las especies contribuyen de forma desigual al almacenamiento de biomasa y carbono en los bosques del Sur y Noroccidente del departamento de Córdoba. Se evidencia que la mayor cantidad de carbono está depositado en la biomasa de un pequeño número de especies, en promedio cinco especies contienen entre el 50 y 80% de las reservas de biomasa y carbono en cada tipo de bosque. Entre las especies que exhiben esta condición se encuentran: *Quercus humboldtii* (62.99% de biomasa en el tipo de bosque donde se encuentra), *Peltogyne purpurea* (37.88%), *Guazuma ulmifolia* (35.16%), *Cavanillesia platanifolia* (34.74%), *Pseudobombax septenatum* (32.09%), Bombacaceae sp. 1 (31.02%), *Dipteryx oleifera* (23.25%), *Pentaclethra macroloba* (22.59%) y *Pentaplaris doroteae* (26.39%). La contribución de las especies al almacenamiento de carbono en estos bosques está determinada principalmente por la talla del árbol (área basal) y en segundo lugar por la abundancia (número de individuos).

Esta condición indica que si se ejerce presión antrópica (tala) sobre las poblaciones de estas espcies se liberará a su vez una cantidad importante de carbono. Según Estupinán & Jiménez (2010), la especie

*Dipteryx oleifera* representa una importante fuente de uso para las comunidades de la región dentro de las especies arbóreas de gran porte. Al analizar su relación con los altos contenidos de carbono en la zona se concluye que su amplio uso en comestibles, construcción, y elaboración de herramientas debe ser controlado debido a que sería causante de un gran aporte de GEI en el territorio.

En este aporte, *Dipteryx oleifera* fue la especie con mayor proporción de biomasa en las asociaciones Jacarando copaiae - Pouterietum multiflorae, Prestoeo decurrentis - Trichillietum poeppigi y Protio aracouchini - Viroletum elongatae con un promedio de 20% de contribución a las reservas de carbono. Si la mayor contribución a las reservas de carbono en los bosques está dada por unas pocas especies, es preciso que el uso que los habitantes de la zona le dan a estas, esté acorde con medidas de mitigación de los impactos ya que pueden situarse en condición vulnerable o peligro por la tala, factor de gran influencia en la permanencia de dichas reservas en el tiempo. Con este resultado se dan argumentos para la conservación de especies, un argumento más es su papel significativo en el mantenimiento de servicios ambientales como el almacenamiento de carbono.

Los bosques las asociaciones Macrolobio ischnocalycis - Peltogynetum purpurea, Prestoeo decurrentis - Trichillietum poeppigi, Tovomito weddellianae - Quercetum humboldtii ubicados en el parque natural nacional Paramillo y de la asociación Protio aracouchini - Viroletum elongatae ubicado en la zona de amortiguación, presentaron los valores más altos de carbono en la biomasa con respecto a los demás bosques lo cual se relaciona con el buen estado de conservación que presentan. Estos resultados sumados a su ubicación en una zona con prioridad de conservación y la disposición de algunos pobladores a participar en planes que tiendan a la protección de los bosques, son condiciones que constituyen una ventaja para proponer estas áreas como estratégicas para la protección de las reservas de carbono y biodiversidad y motivan la búsqueda de incentivos económicos para su mantenimiento como es el caso de proyectos de reducción de emisiones por deforestación evitada y biodiversidad REDD+.

El papel de los bosques con alto grado de intervención y secundarios también es importante, aunque no hayan presentado cantidades altas de biomasa, pueden constituir sumideros de carbono debido a la fase en que se encuentran (recuperación y crecimiento) en la cual capturan dióxido de carbono y lo almacenan en sus estructuras. Además, si estos bosques ocupan áreas extensas del paisaje en Córdoba, que aunque no se ha cuantificado (investigación que está en desarrollo), se prevee que tienen una gran proporción en el espacio ocupado, por consiguiente en conjunto sumarían una cantidad importante de reservas. Estas condiciones pueden hacer de los bosques intervenidos descritos en cuanto a su biomasa y carbono en localidades de Córdoba, tales como los pertenecientes a las asociaciones Jacarando copaiae - Pouterietum multiflorae y Marilo laxiflorae - Pentaclethretum macrolobae jueguen un papel importante en el ciclo del carbono, como lo han descrito en otros estudios (Sierra *et al.* 2007).

## AGRADECIMIENTOS

## LITERATURA CITADA

**Anexo 17.** Biomasa de especies dominantes para cada tipo de bosque.

| Asociación | Especie | Biomasa (t/0.05 ha) | Biomasa (%) | RCB (t/0.05 ha) | RCB (%) |
|---|---|---|---|---|---|
| Acalypha sp. y Guazuma ulmifolia | Guazuma ulmifolia | 2,59 | 35,16 | 1,19 | 35,52 |
| | Tabebuia rosea | 0,81 | 10,96 | 0,38 | 11,47 |
| | Spondias mombin | 0,6 | 8,2 | 0,28 | 8,28 |
| | Caesalpinia glabrata | 0,35 | 4,76 | 0,15 | 4,59 |
| | Mimosaceae sp. 6 | 0,32 | 4,35 | 0,14 | 4,29 |
| | Subtotal | 4,67 | 63,43 | 2,14 | 64,15 |
| | Otras especies | 2,69 | 36,57 | 1,2 | 35,85 |
| | Total | 7,36 | 100 | 3,34 | 100 |
| Cappari odoratissimatis - Cavanillesietum platanifoliae | Bombacaceae sp. 1 | 6,95 | 31,02 | 3,14 | 31,07 |
| | Cavanillesia platanifolia | 5,83 | 26,03 | 2,63 | 25,96 |
| | Ruprechtia costata | 3,15 | 14,05 | 1,43 | 14,13 |
| | Manilkara bidentata | 0,85 | 3,81 | 0,39 | 3,82 |
| | Machaerium milleflorum | 0,76 | 3,39 | 0,34 | 3,39 |
| | Subtotal | 17,54 | 78,3 | 7,93 | 78,37 |
| | Otras especies | 4,85 | 21,7 | 2,19 | 21,63 |
| | Total | 22,39 | 100 | 10,12 | 100 |
| Cariniano pyriformis - Pentaplarietum doroteae | Pentaplaris sp. 1 | 5,3 | 26,39 | 2,33 | 25,78 |
| | Cavanillesia platanifolia | 4,85 | 24,15 | 2,17 | 24,01 |
| | Castilla elastica | 3,1 | 15,43 | 1,43 | 15,82 |
| | Croton sp. 1 | 1,79 | 8,91 | 0,82 | 9,07 |
| | Tetragastris panamensis | 0,68 | 3,39 | 0,32 | 3,54 |
| | Subtotal | 15,72 | 78,27 | 7,07 | 78,22 |
| | Otras especies | 4,37 | 21,73 | 1,97 | 21,78 |
| | Total | 20,09 | 100 | 9,04 | 100 |
| Cordietum proctato - Cordietum panamensis | Pentaplaris sp. 1 | 2,31 | 22,14 | 1,02 | 21,66 |
| | Cavanillesia platanifolia | 1,31 | 12,58 | 0,59 | 12,52 |
| | Bursera simaruba | 0,57 | 5,45 | 0,26 | 5,45 |
| | Acanthaceae sp. 2 | 0,48 | 4,63 | 0,22 | 4,66 |
| | Pterocarpus acapulcensis | 0,44 | 4,25 | 0,2 | 4,31 |
| | Subtotal | 5,11 | 49,05 | 2,29 | 48,6 |
| | Otras especies | 5,3 | 50,95 | 2,41 | 51,4 |
| | Total | 10,41 | 100 | 4,7 | 100 |
| Jacarando copaiae - Pouterietum multiflorae | Dipteryx oleifera | 2,98 | 16,16 | 1,41 | 16,45 |
| | Dialium guianense | 2,44 | 13,22 | 1,14 | 13,23 |
| | Pouteria multiflora | 1,03 | 5,58 | 0,47 | 5,43 |
| | Pentaclethra macroloba | 0,81 | 4,37 | 0,38 | 4,46 |
| | Eschweilera pittieri | 0,77 | 4,19 | 0,37 | 4,34 |
| | Subtotal | 8,03 | 43,52 | 3,77 | 43,91 |
| | Otras especies | 10,43 | 56,48 | 4,82 | 56,09 |
| | Total | 18,46 | 100 | 8,59 | 100 |
| Macrolobio ischnocalycis - Peltogynetum purpurea | Peltogyne purpurea | 15,57 | 37,88 | 7,11 | 37,61 |
| | Couratari guianensis | 4,18 | 10,17 | 1,94 | 10,25 |
| | Andira inermis | 3,93 | 9,57 | 1,8 | 9,52 |
| | Huberodendron patinoi | 3,82 | 9,3 | 1,76 | 9,28 |
| | Licania sp. 2 | 3,19 | 7,76 | 1,47 | 7,75 |
| | Subtotal | 30,69 | 74,68 | 14,08 | 74,41 |
| | Otras especies | 24,09 | 25,32 | 11,12 | 25,59 |
| | Total | 54,78 | 100 | 25,2 | 100 |

**Continuación Anexo 17.**

| Asociación | Especie | Biomasa (t/0.05 ha) | Biomasa (%) | RCB (t/0.05 ha) | RCB (%) |
|---|---|---|---|---|---|
| Marilo laxiflorae - Pentaclethretum macrolobae | Pentaclethra macroloba | 3,18 | 22,59 | 1,5 | 22,82 |
| | Sloanea tuerckheimii | 2,08 | 14,8 | 0,99 | 14,99 |
| | Brosimum utile | 1,68 | 11,95 | 0,79 | 12,02 |
| | Sloanea zuliaensis | 0,98 | 6,97 | 0,46 | 7 |
| | Ermiriodes sp. 1 | 0,64 | 4,55 | 0,3 | 4,49 |
| | Subtotal | 8,56 | 60,86 | 4,04 | 61,32 |
| | Otras especies | 5,5 | 39,14 | 2,53 | 38,68 |
| | Total | 14,06 | 100 | 6,57 | 100 |
| Mayno grandifoliae - Astrocaryetum malybo | Pseudobombax septenatum | 4,4 | 32,09 | 1,94 | 31,19 |
| | Schizolobium parahyba | 1,35 | 9,85 | 0,62 | 9,97 |
| | Astrocaryum malybo | 1,19 | 8,68 | 0,54 | 8,68 |
| | Aegiphila sp. 1 | 1 | 7,29 | 0,46 | 7,4 |
| | Bursera simaruba | 0,95 | 6,93 | 0,43 | 6,91 |
| | Subtotal | 8,89 | 64,84 | 3,99 | 64,15 |
| | Otras especies | 4,82 | 35,16 | 2,23 | 35,85 |
| | Total | 13,71 | 100 | 6,22 | 100 |
| Prestoeo decurrentis - Trichillietum poeppigi | Dipteryx oleifera | 8,21 | 20,04 | 3,77 | 20,14 |
| | Anacardium excelsum | 5,1 | 12,45 | 2,29 | 12,22 |
| | Ceiba pentandra | 2,85 | 6,96 | 1,31 | 6,97 |
| | Vitex cimosa | 2,65 | 6,45 | 1,15 | 6,13 |
| | Chrysophyllum argenteum | 2,19 | 5,34 | 1,01 | 5,37 |
| | Subtotal | 21 | 51,24 | 9,53 | 50,83 |
| | Otras especies | 19,99 | 48,76 | 9,19 | 49,17 |
| | Total | 40,99 | 100 | 18,72 | 100 |
| Protio aracouchini - Viroletum elongatae | Dipteryx oleifera | 7,97 | 23,25 | 3,76 | 23,99 |
| | Virola reidii | 4,43 | 12,92 | 1,92 | 12,24 |
| | Brosimum guianense | 3,32 | 9,7 | 1,44 | 9,18 |
| | Pouteria sp. 5 | 2,17 | 6,33 | 0,98 | 6,24 |
| | Virola sebifera | 1,96 | 5,72 | 0,91 | 5,79 |
| | Subtotal | 19,85 | 57,92 | 9,01 | 57,44 |
| | Otras especies | 14,42 | 42,08 | 6,65 | 42,56 |
| | Total | 34,27 | 100 | 15,66 | 100 |
| Tovomito weddellianae - Quercetum humboldtii | Quercus humboldtii | 18,86 | 62,99 | 8,68 | 62,93 |
| | Aspidosperma sp. 3 | 1,88 | 6,28 | 0,87 | 6,27 |
| | Nectandra martinicensis | 1,54 | 5,14 | 0,71 | 5,11 |
| | Virola elongata | 1,26 | 4,21 | 0,59 | 4,28 |
| | Dendrobangia boliviana | 1,09 | 3,62 | 0,5 | 3,59 |
| | Subtotal | 24,63 | 82,24 | 11,35 | 82,18 |
| | Otras especies | 5,31 | 17,76 | 2,44 | 17,82 |
| | Total | 29,94 | 100 | 13,79 | 100 |
| Trichilio hirtae- Schizolobietum parahibi | Cavanillesia platanifolia | 8,36 | 34,74 | 3,78 | 34,88 |
| | Trichilia hirta | 3,18 | 13,2 | 1,44 | 13,3 |
| | Pentaplaris doroteae | 2,37 | 9,87 | 1,04 | 9,64 |
| | Pseudobombax septenatum | 1,28 | 5,33 | 0,56 | 5,2 |
| | Bursera simaruba | 1,1 | 4,57 | 0,5 | 4,59 |
| | Subtotal | 16,29 | 67,71 | 7,32 | 67,62 |
| | Otras especies | 7,77 | 32,29 | 3,51 | 32,38 |
| | Total | 24,06 | 100 | 10,82 | 100 |